\documentclass[journal, onecolumn, 12pt]{IEEEtran}%单栏，字体为12pt
\usepackage{tikz}
\usetikzlibrary{arrows,shapes,chains,positioning}
\usepackage{diagbox}
\usepackage{amsmath,amssymb,amsfonts,slashbox}
\usepackage{amsmath}
\usepackage[ruled,vlined]{algorithm2e}
\usepackage{epstopdf}
\usepackage{graphicx}
\usepackage{subfigure}
\usepackage[percent]{overpic}
\usepackage{float}
\usepackage{booktabs}
\usepackage{multirow}
\usepackage{tabularx}
\usepackage{capt-of}
\usepackage{ulem}
\usepackage{bm}
\usepackage{cite}
\usepackage{setspace}
\usepackage{lineno}
\usepackage{setspace}
\usepackage{authblk}
\def\BibTeX{{\rm B\kern-.05.em{\sc i\kern-.025em b}\kern-.08em
 T\kern-.1667em\lower.7ex\hbox{E}\kern-.125emX}}
\begin{document}

\title{Vector Approximate Message Passing based Channel Estimation for MIMO-OFDM Underwater Acoustic Communications}

\author{\normalsize Author List}
\author{\normalsize Wenxuan Chen, Jun Tao,~\IEEEmembership{Senior Member,~IEEE,} Lu Ma,~\IEEEmembership{Member,~IEEE,} \\and Gang Qiao,~\IEEEmembership{Member,~IEEE}
\thanks{W. Chen and J. Tao are with the Key Laboratory of Underwater Acoustic Signal Processing of the Ministry of Education, School of Information Science and Engineering, Southeast University, Nanjing, 210096, China. L. Ma and G. Qiao are with the Acoustic Science and Technology Laboratory, the Key Laboratory of Marine Information Acquisition and Security, Ministry of Industry and Information Technology, and the College of Underwater Acoustic Engineering, Harbin Engineering University, Harbin, 150001, China. (Email: jtao@seu.edu.cn)}
}
%\thanks{W. Chen and J. Tao is with the Key Laboratory of Underwater Acoustic Signal Processing of the Ministry of Education, \\
%School of Information Science and Engineering, Southeast University, Nanjing, 210096, China (Email: jtao@seu.edu.cn).}
%%the Acoustic Science and Technology Laboratory, Harbin Engineering Univ., Harbin, 150001, China

\maketitle
\graphicspath{{./epsfigure/}}

\begin{abstract}
Accurate channel estimation is critical to the performance of orthogonal frequency-division multiplexing (OFDM) underwater acoustic (UWA) communications, especially under multiple-input multiple-output (MIMO) scenarios. In this paper, we explore Vector Approximate Message Passing (VAMP) coupled with Expected Maximum (EM) to obtain channel estimation (CE) for MIMO OFDM UWA communications. The EM-VAMP-CE scheme is developed by employing a Bernoulli-Gaussian (BG) prior distribution for the channel impulse response, and hyperparameters of the BG prior distribution are learned via the EM algorithm. Performance of the EM-VAMP-CE is evaluated through both synthesized data and real data collected in two at-sea UWA communication experiments.
It is shown the EM-VAMP-CE achieves better performance-complexity tradeoff compared with existing channel estimation methods.
\end{abstract}

\begin{IEEEkeywords}
Channel Estimation (CE), Multiple-Input Multiple-Output (MIMO), Orthogonal Frequency-Division Multiplexing (OFDM), Underwater Acoustic (UWA) Communications, Vector Approximate Message Passing (VAMP).
\end{IEEEkeywords}
%Multiple-Input Multiple-Output (MIMO), Expected Maximum (EM),

\section{\label{sec:1} Introduction}

Due to its limited bandwidth, multipath fading, and significant Doppler effect, the underwater acoustic (UWA) channel has long been recognized as one of the most challenging communication media \cite{Tao18}. Attributed to the progress in Doppler compensation, orthogonal frequency-division multiplexing (OFDM) modulation is receiving many attentions in UWA communication, for its immunity to multipath propagation, high spectral efficiency, and so on. The channel estimation (CE) is one of key factors in the performance of OFDM UWA communications \cite{Barazideh19,Qiao19}.

Conventional least squares (LS) and linear minimum mean square error (LMMSE) channel estimations are found in OFDM UWA communications \cite{Jingjing08,Zhang08,Huangxiao10,Yonggang11,Tao18}. Even though, an underwater acoustic channel generally has a long delay spread, making aforementioned traditional methods inefficient in the sense that a large pilot overhead is required \cite{Zhang08,Tao18}.
UWA channels usually manifest sparse property, meaning that a large percentage of channel energy is concentrated at a small amount of taps. This observation naturally motivates the development of sparsity-aware channel estimation so as to reduce pilot overhead and/or complexity. Initial effort was found in single-carrier UWA communications \cite{Li07}, where \textcolor{black}{compressive sensing (CS) algorithms including matching pursuit (MP), orthogonal matching pursuit (OMP), etc. were employed to achieve channel estimation.} The development of sparsity-aware channel estimations for OFDM UWA communications can be found in \cite{Yu15,Chen17,Wang21,Huang10,Yin15}, again mainly attributed to the CS techniques. In \cite{Yu15,Chen17,Wang21}, the greedy OMP algorithm and its variants were used. In \cite{Yu15}, block fast Fourier transform (FFT) was adopted to reduce the implementation complexity of OMP channel estimation. In \cite{Chen17}, the OMP algorithm was employed for joint channel estimation and impulsive noise mitigation in OFDM underwater acoustic communications.
In \cite{Wang21}, channel estimation for OFDM UWA communications was achieved via adaptive OMP (A-OMP) algorithm, which outperforms standard OMP method for improved termination condition. In \cite{Huang10,Yin15}, convex-optimization-type CS methods including Basis Pursuit (BP) and Basis Pursuit denoising (BPDN) were adopted. In \cite{Huang10}, three implementations of BP, named the l1$\underline{~}$ls, SpaRSA, and YALL1, were used for estimating time-varying sparse channels. Simulation and experimental results showed they were comparable in performance, but the SpaRSA and YALL1 had lower complexity. In \cite{Yin15}, a BPDN-based channel estimation was proposed for OFDM UWA communications employing virtual time reversal processing. It slightly outperformed the channel estimation via MP algorithm. In \cite{Berger10}, channel estimations based on subspace methods ESPRIT and Root-MUSIC as well as CS algorithms including OMP and BP, were investigated and compared for OFDM UWA communications under different Doppler conditions. A path-based channel model was adopted, that is the channel is described by a limited number of paths, each characterized by a delay, Doppler scale, and attenuation factor. It showed the CS based CE has better performance than the subspace based CE, and the BP-CE slightly outperforms the OMP-CE. %and

Bayesian inference techniques like the maximum a posteriori probability (MAP) estimation, sparse Bayesian learning (SBL) algorithm \cite{Tipping01,Wipf04}, and approximate message passing (AMP) techniques \cite{Donoho09,Bayati11} and variants, have found success in terrestrial OFDM systems \cite{Kim08,Prasad14,Wu16}. Due to the improved performance and robustness, they have also attracted many attentions in OFDM UWA systems \cite{Panayirci19,Qiao18,Wang20,Wu19} recently.
In \cite{Panayirci19}, an OMP-MAP algorithm was proposed for OFDM UWA channel estimation. With the path delays and Doppler spread identified via OMP, path gains were estimated using the MAP technique. The OMP-MAP channel estimation outperforms conventional OMP CE. In \cite{Qiao18}, an SBL based channel estimation scheme was proposed for OFDM UWA communications. To further improve the performance, temporal correlation in channels across consecutive OFDM blocks was utilized, leading to the so-called temporal multiple SBL (TMSBL) channel estimator. In \cite{Wang20}, the SBL was employed for the joint estimation and tracking of channel and impulsive noise in OFDM UWA systems. Compared with traditional SBL estimation, this work exploited the joint sparsity of channel and impulse noise for better estimation performance.

Despite decent performance, the SBL-CE and MAP-CE often suffer high computation complexity, which limits their practical applications \cite{Lee17,Lv19}. In contrast, the AMP-type techniques have the advantage of lower complexity \cite{Shoukairi18,Zhu19}. The generalized AMP (GAMP) \cite{Rangan11} as a generalization of AMP \cite{Donoho09}, has been used for channel estimation in OFDM UWA communications suffering impulsive noise \cite{Wu19}. The GAMP, however, works well only when the measurement matrix consists of independent and identically distributed (i.i.d.) Gaussian elements. Therefore, its practical application is limited.  %\textcolor{blue}{Combined with EM algorithm, it achieves decent performance.}
Recently, the vector AMP (VAMP) algorithm was proposed \cite{Rangan17,Rangan19}. It is applicable to a wider class of right-rotationally invariant measurement matrices. In light of this, we have made a preliminary investigation on the feasibility of VAMP based channel estimation for OFDM UWA communication systems \cite{Chen21}. In this paper, a much more comprehensive study on the VAMP based channel estimation was presented for multiple-input multiple-output (MIMO) OFDM UWA communication system. The Bernoulli-Gaussian (BG) prior distribution is adopted for the UWA channel to account for its sparsity. Parameters of the BG distribution are learned from the received data via the Expectation Maximization (EM) algorithm, instead of being set empirically as in \cite{Chen21}. The complexity of the resulting EM-VAMP-CE was analyzed and compared with existing LS, LMMSE, OMP, and SBL CEs, under scenarios of non-orthogonal pilot sequences and orthogonal pilot sequences. It showed the complexity of the EM-VAMP-CE is significantly lower than that of the SBL-CE when non-orthogonal pilot sequences were employed. To verify the performance of the EM-VAMP-CE, MIMO OFDM UWA communication systems employing overlapping pilot pattern across transducers \cite{Tao18} were considered. Both synthesized data and real data were used for verification. In the simulation, orthogonal pilot sequences \cite{Barhumi03} were adopted. The real data were collected in two UWA communication experiments: XM16 and SPACE08, where non-orthogonal pilot sequences were adopted. For comparison purpose, channel estimations based on LS, LMMSE, OMP, and SBL were also included. Both simulation and experimental results show the proposed EM-VAMP-CE achieves the best performance-complexity tradeoff at a low pilot overhead. %The performance of the proposed EM-VAMP-CE scheme is evaluated by real data collected during the in terms of the bit error rate (BER). for the involvement of matrix inversion  with a high transmission efficiency

The rest of this paper is organized as follows. In Section~\ref{MIMOsystemmodel}, the system model for zero-padding (ZP) MIMO OFDM UWA communications is presented. In Section~\ref{VAMPestimator}, the EM-VAMP based channel estimator is derived. Section~\ref{SimulationResult} and Section~\ref{ExperimentResult} present simulation and experimental results, repectively. Section~\ref{Conclusion} concludes this paper. %discusses performance of the EM-VAMP estimator comparing with other conventional estimators.

{\bf Notation}: The ${\mathcal{C}}^{p\times q}$ represents a complex space of dimension $p\times q$. The superscripts $(\cdot)^T$ and $(\cdot)^H$ represent the transpose and Hermitian, respectively. The $\odot$ denotes the Hadamard product. The $\left\langle{\cdot}\right\rangle$ is an empirical averaging operation defined as $\left\langle{{\bf x}}\right\rangle = \frac{1}{P}\sum_{p=1}^{P}{x_p}$, where $x_p$ is the $p$-th element of the size-$P$ vector ${\bf x}$. The $\text{Diag}({\bf a})$ represents a diagonal matrix with the vector {\bf a} on its diagonal. The ${\mathbb{E}}[\cdot]$ denotes an expectation operation and the $\text{Tr}[\cdot]$ takes the trace of a matrix. The $\|\cdot\|_{F}$ represents the Frobenius norm, and the ${\bf I}$ is an identity matrix. The ${\mathcal C}{\mathcal N(x;\mu,{\sigma}^2)}{\triangleq}{\frac{1}{\pi{\sigma}^2}\text{exp}(\frac{-{\vert{x-\mu}\vert}^2}{{\sigma}^2})}$ is the probability density function of a circularly symmetric complex Gaussian distribution with mean $\mu$ and variance ${\sigma}^2$. The ${\bf F}_K$ denotes a size-$K$ normalized DFT matrix with its $(m,n)$-th element being $w_{m,n}=\frac{1}{\sqrt{K}}e^{-j2\pi\frac{mn}{K}}$ for $m,n=0,1\cdots,K-1$. The $\delta(\cdot)$ is a Dirac delta function.

\section{\label{MIMOsystemmodel} System Model}

An MIMO OFDM UWA communication system with $N$ transducers and $M$ hydrophones is considered. To maintain a high transmission efficiency, the same pilot subcarrier index set, ${\mathcal P} =[ I_1, I_2, \cdots, I_{K_p}]$, is employed across all transducers. In addition to the $K_p$ pilot symbols, each OFDM block carries $K_d$ data symbols, resulting in a block length of $K = K_p + K_d$\cite{Tao18}. On the $n$-th transducer, a block of pilot/data symbols ${\bf X}_n = [X_n(0), X_n(1), \cdots ,X_n(K-1)]^T$ are modulated onto different subcarriers through an inverse discrete Fourier transform (IDFT) operation, leading to the time-domain OFDM symbol ${\bf x}_n = {\bf F}_K^H{\bf X}_n = [{x}_n(0), {x}_n(1), \cdots ,{x}_n(K-1)]^T$. In order to eliminate the inter-block interference (IBI), a guard interval in form of zero-padding is added at the end of ${\bf x}_n$. The length of the guard interval is chosen as $N_{ZP}\geq L$, with $L$ being the channel length \cite{Muquet02}.

On the receiver side, Doppler preprocessing via resampling and carrier frequency offset (CFO) compensation is first performed, such that the channel can be treated time-invariant over one OFDM block \cite{Tao18}. On the $m$-th hydrophone, the received signal at time $k$ is then represented as
%%
%shift in an UWA channel will destroy the orthogonal property of OFDM signal and inter-carrier interference (ICI)
%is induced, which can be handled by. After that, the channel is assumed to be time-invariant over one OFDM block.
%%
\begin{eqnarray}
    \label{discrete expression}
     \uline{y}_m(k) = \sum_{n=1}^{N}\sum_{l=0}^{L-1}\uline{h}_{m,n}(l){x}_n(k-l)+\uline{n}_m(k)
\end{eqnarray}
where $\uline{h}_{m,n}(l)$ is the $l$-th tap of the discrete-time subchannel between the $n$-th transducer and the $m$-th
hydrophone. After performing an overlap-adding (OLA) operation as follows \cite{Muquet02}
\begin{eqnarray}
\begin{aligned}
    \label{overlap-adding}
     \uline{\bf y}_m = &[\uline{y}_m(0), \uline{y}_m(1), \uline{y}_m(2), \cdots, \uline{y}_m(K-1)]\\
      & + [\uline{y}_m(K), \cdots, \uline{y}_m(K+L-2),0 \cdots, 0]
\end{aligned}
\end{eqnarray}
and a $K$-point DFT operation, the frequency-domain vector is obtained as $\tilde{\bf y}_m = {\bf F}_K\uline{\bf y}_m$. For the purpose of channel estimation, the frequency-domain samples corresponding to the $K_p$ pilot subcarriers are extracted out of $\tilde{\bf y}_m$, leading to ${\bf y}_m=[y_{m,I_1}, y_{m,I_2},\cdots, y_{m,I_p}]\in{\mathcal{C}}^{K_p\times 1}$. The ${\bf y}_m$ can be expressed as %corresponding to $K_p$ pilots are expressed by
%Frequency domain vector ${\bf Y}$ can be obtained by applying a $K$-point DFT on
\begin{eqnarray}
\begin{aligned}
    \label{MIMO_system_model_1}
    {\bf y}_m= \sum_{n=1}^{N}{\bf S}_n{\bf F}{\uline{\bf h}}_{m,n}+{\bf n}_m
               = {\bf W}{\uline{\bf h}}_m+{\bf n}_m\\%{\bm\hbar}_m
%               &=[{\bf S}_1,{\bf S}_2,\cdots,{\bf S}_N]
%               \left[\begin{array} {ccc}
%    {\bf F} & {\bf 0} & {\bf 0}\\
%    {\bf 0} & \ddots & {\bf 0}\\
%    {\bf 0} & {\bf 0} & {\bf F}
%    \end{array}\right]{\uline{\bf h}}_m+{\bf n}_m
\end{aligned}
\end{eqnarray}
where ${\bf S}_n = \text{Diag}(X_n(I_1), X_n(I_2), \cdots, X_n(I_{K_p}))$ is a diagonal matrix consisting of $K_p$ pilot symbols sent by the $n$-th transducer, ${\bf F}\in{\mathcal{C}}^{K_p\times L}$ is made up of $K_p$ rows with indices given in $\mathcal P$ and the first $L$ columns of ${\bf F}_K$, ${\bf W}=[{\bf S}_1{\bf F},{\bf S}_2{\bf F},\cdots,{\bf S}_N{\bf F}]\in{\mathcal{C}}^{K_p\times NL}$. The ${\uline{\bf h}}_{m,n}=[\uline{h}_{m,n}(0),\uline{h}_{m,n}(1),\cdots,\uline{h}_{m,n}(L-1)]^T\in{\mathcal{C}}^{L\times 1}$, and ${\uline{\bf h}}_m=[{\uline{\bf h}}^h_{m,1},{\uline{\bf h}}^h_{m,2},\cdots,{\uline{\bf h}}^h_{m,N}]^h\in{\mathcal{C}}^{NL\times 1}$. The ${\bf n}_m\in{\mathcal{C}}^{K_p\times 1}$ is the frequency-domain noise vector, assumed to follow a complex Gaussian distribution ${\mathcal {CN}({\bf 0},{\gamma_w}^{-1}{\bf I})}$. It is noted when a path-based channel model \cite{Berger10} is considered, the system model in \eqref{MIMO_system_model_1} is still valid if only the Doppler effect can be properly compensated. The goal of channel estimation is to estimate ${\uline{\bf h}}_m$ based on \eqref{MIMO_system_model_1}.

\section{EM-VAMP based channel estimation\label{VAMPestimator}}

The problem given by \eqref{MIMO_system_model_1} is a standard linear regression (SLR) problem, which can be solved via the VAMP algorithm. In the following, we first present the VAMP based channel estimation, temporarily assuming model parameters and super-parameters are known. After that, parameter estimation via EM method is introduced. Last, a complexity comparison between the proposed EM-VAMP channel estimation scheme and existing methods is made.

\subsection{\label{VAMP} VAMP based channel estimation}

%Due to the advantage of VAMP in solving SLR problem, it is chosen for channel estimation.
The VAMP algorithm provides an iterative procedure for approximately solving the MAP or minimum mean-squared error (MMSE) estimation of ${\uline{\bf h}}_m$ based on observed data $ {\bf y}_m$ in \eqref{MIMO_system_model_1}. In this paper, we seek an MMSE estimation of ${\uline{\bf h}}_m$.

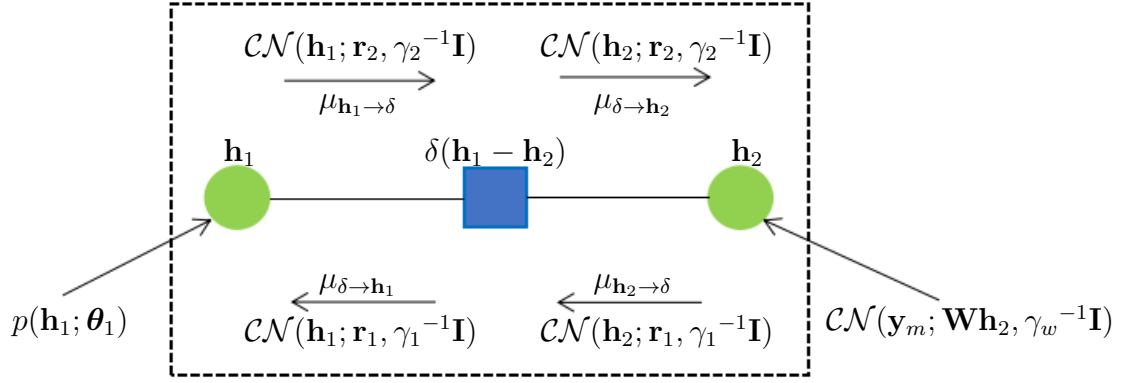
\begin{figure}[!ht]
\centering
%\scriptsize
\begin{overpic}[scale=0.8]{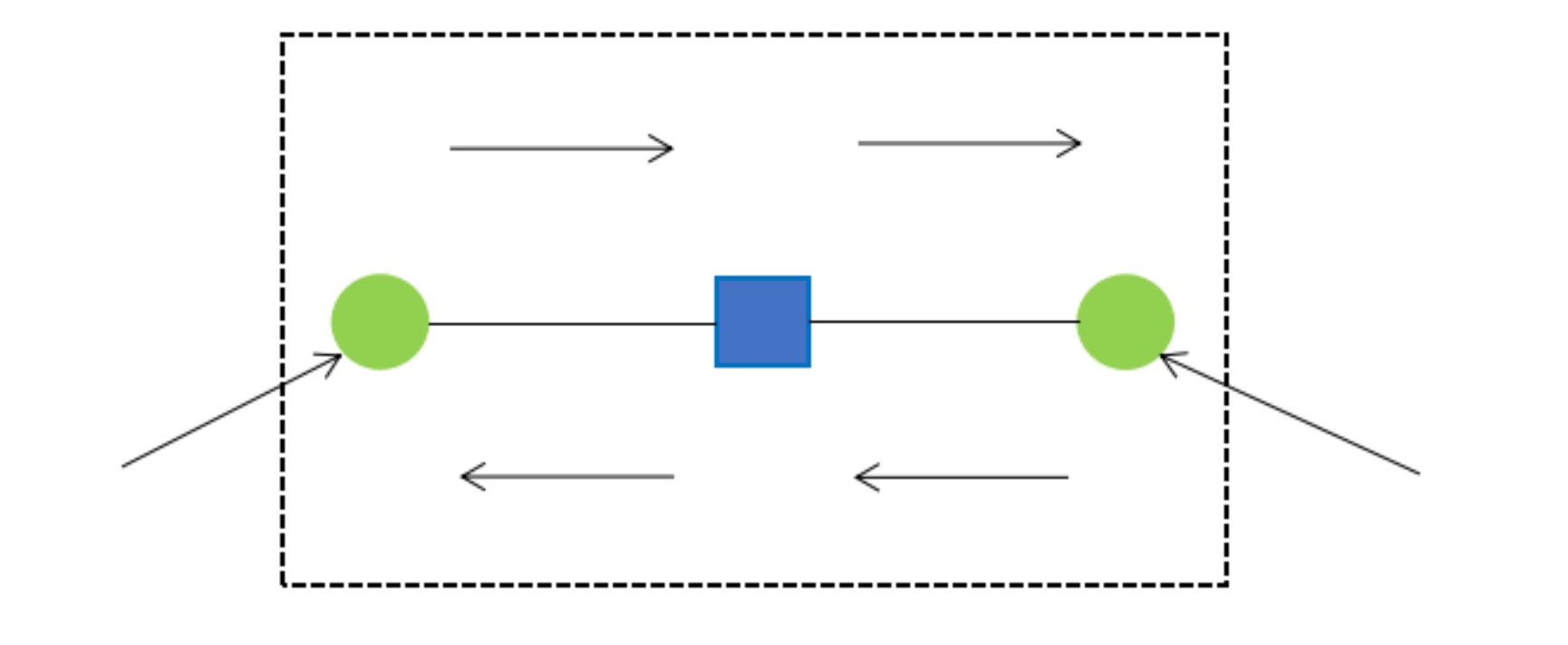}
\put(3,8){$ p({\bf h}_1;{\bm \theta}_1)$}
%\put(26,25){${\bf h}_1\sim $}
%\put(15,21){$\mathcal CN({\bf h}_1;{\bf{\widehat h}}_1,{\eta_1}^{-1}{\bf I})$}
\put(23,24){${\bf h}_1$}
\put(42,24){$\delta({\bf h}_1-{\bf h}_2)$}
%\put(66,25){$ {\bf h}_2\sim$}
%\put(58,21){$\mathcal CN({\bf h}_2;{\bf{\widehat h}}_2,{\eta_2}^{-1}{\bf I})$}
\put(71,24){$ {\bf h}_2$}
\put(80,8){$ \mathcal {CN}({\bf y}_m;{\bf W}{\bf h}_2,{\gamma_w}^{-1}{\bf I})$}
\put(25,7){$ \mathcal {CN}({\bf h}_1;{\bf r}_1,{\gamma_1}^{-1}{\bf I})$}
\put(32,12){$\mu_{\delta\rightarrow{\bf h}_1}$}
\put(53,7){$\mathcal {CN}({\bf h}_2;{\bf r}_1,{\gamma_1}^{-1}{\bf I})$}
\put(58,12){$\mu_{{\bf h}_2\rightarrow\delta}$}
\put(25,34){$ \mathcal {CN}({\bf h}_1;{\bf r}_2,{\gamma_2}^{-1}{\bf I})$}
\put(32,29){$\mu_{{\bf h}_1\rightarrow\delta}$}
\put(53,34){$\mathcal {CN}({\bf h}_2;{\bf r}_2,{\gamma_2}^{-1}{\bf I})$}
\put(58,29){$\mu_{\delta\rightarrow{\bf h}_2}$}
\end{overpic}
\caption{The factor graph for VAMP channel estimation (the circle and square nodes represent, respectively, a variable node and a factor node).}\label{VAMP_factor}
\end{figure}

We assume ${\uline{\bf h}}_m$ has i.i.d elements and to characterize its sparsity, each element follows a Bernoulli-Gaussian prior distribution. Therefore
\begin{eqnarray}
\begin{aligned}
    \label{VAMPestimator_1}
     p({\uline{\bf h}}_m;{\bm \theta}_1) &= {\prod_{i=1}^{NL}} p({\uline h}_m(i);{\bm \theta}_1)\\
     &= {\prod_{i=1}^{NL}}(1-\lambda)\delta({\uline h}_m(i))+\lambda{\mathcal {CN}({\uline h}_m(i);0,{\gamma_h}^{-1})}
\end{aligned}
\end{eqnarray}
where ${\bm \theta}_1 = [\lambda, \gamma_h]^{T}$ contains the hyper-parameters, $\lambda \in (0,1)$ is the normalized sparsity of the channel, and $\gamma_h$ denotes the precision (inverse variance) of the distribution of a non-zero channel element. From \eqref{MIMO_system_model_1}, it is easy to have
\begin{eqnarray}
\begin{aligned}
    \label{eqn_23sept21_1}
    p({\bf y}_m,{\uline{\bf h}}_m;{\bm \theta}) &= p({\uline{\bf h}}_m;{\bm \theta}_1)p({\bf y}_m|{\uline{\bf h}}_m;{\bm \theta})\\
    &=p({\uline{\bf h}}_m;{\bm \theta}_1)\mathcal {CN}({\bf y}_m;{\bf W}{\uline{\bf h}}_m,{\gamma_w}^{-1}{\bf I})
\end{aligned}
\end{eqnarray}
%, and the likelihood function $p({\bf y}_m|{\uline{\bf h}}_m;{\bm \theta})={\mathcal {CN}({\bf y}_m;{\bf W}{\uline{\bf h}}_m,{\gamma_w}^{-1}{\bf I})}$ has been used
where ${\bm \theta}=[{\bm \theta}_1^T~{\gamma_w}]^T$. Splitting ${\uline{\bf h}}_m$ into two identical variables ${\bf h}_1={\bf h}_2$ \cite{Rangan19}, leads to the following equivalent factorization
%To present the development of VAMP-CE, we start with the following
\begin{eqnarray}
    \label{VAMPestimator_3}
    p({\bf y}_m,{\bf h}_1,{\bf h}_2;{\bm \theta}) = p({\bf h}_1;{\bm \theta}_1)\delta({\bf h}_1-{\bf h}_2){\mathcal {CN}({\bf y}_m;{\bf W}{\bf h}_2,{\gamma_w}^{-1}{\bf I})}
\end{eqnarray}
A factor graph representation of \eqref{VAMPestimator_3} is given in Fig.~\ref{VAMP_factor}, where we have two variable nodes (VNs) and one factor node (FN).

%As the random noise ${\bf N}_m$ in \eqref{MIMO_system_model_1} is Gaussian, the likelihood function of ${\bf h}_2$ is written as
%\begin{eqnarray}
%    \label{VAMPestimator_2}
%     p({\bf Y}_m|{\bf h}_2;\gamma_w) = {\mathcal {CN}({\bf Y}_m;{\bf W}{\bf h}_2,{\gamma_w}^{-1}{\bf I})}
%\end{eqnarray}

It is well known the exact MMSE estimation of ${\uline{\bf h}}_m$ is $\hat{\uline{\bf h}}_m^{\text{MMSE}}=\mathbb{E}[{\uline{\bf h}}_m|{\bf y}_m]$, which boils down to computing the posterior distribution $p({\uline{\bf h}}_m|{\bf y}_m)$. %which can be achieved by running exact message passing over the factor graph in Fig.~\ref{VAMP_factor}. An exact message passing, however, is intractable.
The VAMP provides an iterative way to approximately compute the posterior distribution, as discussed in the following. %In this paper, we employ the VAMP to seek an approximating MMSE estimation of ${\bf h}_m$.

%\textcolor{blue}{The VAMP algorithm runs in a two-part corresponding to ${\bf h}_1$ and ${\bf h}_2$ iterative fashion. Each part can be divided into two steps: (1)the posteriori estimation and (2)the extraction of message to be transmitted to the other part\cite{Dong19,Dong21}. Information interaction between two parts finally reached a stable state. In the last iteration, the posterior mean of ${\bf h}_1$ is output as the estimation of ${\bf h}$.}
%We next present the $k$-th iteration for the VAMP-based channel estimation.
In the $k$-th iteration of the VAMP-based channel estimation, the sum-product (SP) belief \cite{Rangan19} (posterior probability) of the VN ${\bf h}_1$ can be computed with the channel prior probability $p({\bf h}_1;{\bm \theta}_1)$ and the message $\mu_{\delta\rightarrow{\bf h}_1}$ in form of ${\mathcal {CN}({\bf h}_1;{\bf r}_{1,k},{\gamma_{1,k}}^{-1}{\bf I})}$ from the FN $\delta$. Attributed to the i.i.d assumption, it is equivalent to compute the SP belief for each of its elements, that is
\begin{align}
%     &b_1(h_1(i)|r_{1,k}(i),\gamma_{1,k},{\bm \theta}_1) = \frac{\mathcal {CN}({\bf h}_1;{\bf r}_{1,k},{\gamma_{1,k}}^{-1}{\bf I})p({\bf h}_1;{\bm \theta}_1)}{\int{\mathcal {CN}({\bf h}_1;{\bf r}_{1,k},{\gamma_{1,k}}^{-1}{\bf I})} p({\bf h}_1;{\bm \theta}_1)d{\bf h}_1}\label{VAMPestimator_4}\\
%     &= (1-\pi_k(i)){\delta(h_1(i))}+{\pi_k(i)}{\mathcal {CN}({h_1(i);\mu_k(i),{\nu_k}})}\label{VAMPestimator_5}
b_1(h_1(i)|r_{1,k}(i),\gamma_{1,k},{\bm \theta}_1) &= \frac{\mathcal {CN}(h_1(i);r_{1,k}(i),\gamma_{1,k}^{-1})p(h_1(i);{\bm \theta}_1)}{\int\mathcal {CN}(h_1(i);r_{1,k}(i),{\gamma_{1,k}}^{-1})p(h_1(i);{\bm \theta}_1)dh_1(i)}\label{VAMPestimator_4} \\
     &= (1-\pi_k(i)){\delta(h_1(i))}+{\pi_k(i)}{\mathcal {CN}({h_1(i);\mu_k(i),{\nu_k}})}\label{VAMPestimator_5}
\end{align}
where
\begin{eqnarray}
    \label{VAMPestimator_6}
     \pi_k(i) = \left(1+\frac{(1-\lambda)\mathcal {CN}(0;{r_{1,k}(i)},{\gamma_{1,k}}^{-1})}{\lambda\mathcal {CN}(0;{r_{1,k}(i)},({\gamma_{1,k}}^{-1}+{\gamma_h}^{-1}))}\right)^{-1}
\end{eqnarray}
\begin{eqnarray}
    \label{VAMPestimator_7}
     \mu_k(i) = \frac{\gamma_{1,k}r_{1,k}(i)}{\gamma_{1,k}+\gamma_h}
\end{eqnarray}
\begin{eqnarray}
    \label{VAMPestimator_8}
     \nu_k = \frac{1}{\gamma_{1,k}+\gamma_h}
\end{eqnarray}
Details are referred to Appendix A. The posteriori mean-vector ${\widehat{\bf h}}_{1,k}$ and inverse-variance ${\eta_{1,k}}$ of ${\bf h}_1$ are then obtained as
\begin{eqnarray}
    \label{VAMPestimator_9}
    {\widehat{\bf h}}_{1,k} =  {\bm \pi}_k\odot{\bm \mu}_k = {\bm \pi}_k\odot\frac{\gamma_{1,k}{\bf r}_{1,k}}{\gamma_{1,k}+\gamma_h}
\end{eqnarray}
\begin{eqnarray}
    \label{VAMPestimator_11}
     {\eta_{1,k}} = \frac{\gamma_{1,k}}{\alpha_{1,k}}
\end{eqnarray}
where ${\bm \pi}_k=[\pi_k(1),\pi_k(2),\cdots,\pi_k(NL)]^T$, ${\bm \mu}_k=[\mu_k(1),\mu_k(2),\cdots,\mu_k(NL)]^T$, and
\begin{eqnarray}
    \label{VAMPestimator_10}
     {\alpha_{1,k}} = 	\left\langle{\bm \pi_k}\frac{\gamma_{1,k}}{\gamma_{1,k}+\gamma_h}\right\rangle
\end{eqnarray}

The SP belief of ${\bf h}_1$ is approximated as a complex Gaussian distribution ${\mathcal {CN}({\bf h}_1;{\widehat{\bf h}}_{1,k},{\eta_{1,k}}^{-1}{\bf I})}$, according to the VAMP algorithm. The message to be passed from the VN ${\bf h}_1$ to the FN $\delta$, $\mu_{{\bf h}_1\rightarrow\delta}$, can then be obtained via Gaussian division as
%Based on the message passing rules of expectation propagation(EP), the prior probability of ${\bf h}_2$ is obtained
$\frac{{\mathcal {CN}({\bf h}_1;{\widehat{\bf h}}_{1,k},{\eta_{1,k}}^{-1}{\bf I})}}{{\mathcal {CN}({\bf h}_1;{\bf r}_{1,k},{\gamma_{1,k}}^{-1}{\bf I})}}\propto {\mathcal {CN}({\bf h}_1;{\bf r}_{2,k},{\gamma_{2,k}}^{-1}{\bf I})}$, where
\begin{eqnarray}
    \label{VAMPestimator_12}
     {\bf r}_{2,k} = \frac{({\bf{\widehat h}}_{1,k}-\alpha_{1,k}{\bf r}_{1,k})}{1-\alpha_{1,k}}
\end{eqnarray}
\begin{eqnarray}
    \label{VAMPestimator_13}
     \gamma_{2,k} = \gamma_{1,k}{\frac{1-\alpha_{1,k}}{\alpha_{1,k}}}
\end{eqnarray}

The message $\mu_{{\bf h}_1\rightarrow\delta}$ is forwarded by the FN $\delta$ to the VN ${\bf h}_{2}$ without change, such that $\mu_{\delta\rightarrow{\bf h}_2} = {\mathcal {CN}({\bf h}_2;{\bf r}_{2,k},{\gamma_{2,k}}^{-1}{\bf I})}$.
%$\mu_{\delta\rightarrow{\bf h}_2}=\mu_{{\bf h}_1\rightarrow\delta}={\mathcal {CN}({\bf h}_2;{\bf r}_{2,k},{\gamma_{2,k}}^{-1}{\bf I})}$.
With the message $\mu_{\delta\rightarrow{\bf h}_2}$ and the likelihood function ${\mathcal {CN}({\bf y}_m;{\bf W}{\bf h}_2,{\gamma_w}^{-1}{\bf I})}$, the SP belief of ${\bf h}_2$ can be evaluated as follows
\begin{eqnarray}
    \label{VAMPestimator_14}
      b_2({\bf h}_2|{\bf r}_{2,k},\gamma_{2,k},\gamma_w)= \frac{\mathcal {CN}({\bf h}_2;{\bf r}_{2,k},{\gamma_{2,k}}^{-1}{\bf I})p({\bf y}_m|{\bf h}_2;\gamma_w)}{\int{\mathcal {CN}({\bf h}_2;{\bf r}_{2,k},{\gamma_{2,k}}^{-1}{\bf I})}p({\bf y}_m|{\bf h}_2;\gamma_w)d{\bf h}_2}
\end{eqnarray}
which is Gaussian with its mean ${\widehat{\bf h}}_{2,k}$ and covariance matrix ${\bf D}_k$ given by
\begin{eqnarray}
    \label{VAMPestimator_15}
     {\hat{\bf h}}_{2,k} = (\gamma_w{\bf W}^{H}{\bf W}+{\gamma_{2,k}}{\bf I})^{-1}(\gamma_w{\bf W}^{H}{\bf y}_m+{\gamma_{2,k}}{\bf r}_{2,k})
\end{eqnarray}
\begin{eqnarray}
    \label{VAMPestimator_16}
     {\bf D}_k = (\gamma_w{\bf W}^{H}{\bf W}+{\gamma_{2,k}}{\bf I})^{-1}
\end{eqnarray}
The SP belief of ${\bf h}_2$ is further approximated as complex Gaussian ${\mathcal {CN}({\bf h}_2;{\hat{\bf h}}_{2,k},{\eta_{2,k}}^{-1}{\bf I})}$, where
\begin{eqnarray}
    \label{VAMPestimator_17}
      \eta_{2,k} = \frac{\gamma_{2,k}}{\alpha_{2,k}}
\end{eqnarray}
\begin{eqnarray}
    \label{VAMPestimator_18}
     \alpha_{2,k} = {\frac{\gamma_{2,k}}{N}}\text {Tr}{[{\bf D}_k]}
\end{eqnarray}
In \eqref{VAMPestimator_15} and \eqref{VAMPestimator_16}, matrix inversion is required. When orthogonal pilot sequences \cite{Barhumi03} are employed, ${\bf W}^{H}{\bf W}$ is diagonal so the matrix inversion only incurs a low complexity. Otherwise, ${\bf W}$ can be replaced by its  singular value decomposition (SVD) ${\bf W} = {\bf U}{\bf S}{\bf V}^{H}$ such that \eqref{VAMPestimator_15} and \eqref{VAMPestimator_18} become
\begin{eqnarray}
    \label{VAMPestimator_19}
     {\widehat{\bf h}}_{2,k} = {\bf V}{\bf d}_k({\widetilde {\bf y}}_m + \gamma_{2,k}{\bf V}^H{\bf r}_{2,k})
\end{eqnarray}
\begin{eqnarray}
    \label{VAMPestimator_20}
     \alpha_{2,k} = \frac{1}{R}\sum_{n}\frac{\gamma_{2,k}}{\gamma_w{s}_n^2+\gamma_{2,k}},
\end{eqnarray}
where ${R} = \text{rank}({\bf W})$, ${s}_n = [{\bf S}]_{nn}$, and
\begin{eqnarray}
    \label{VAMPestimator_21}
     {\bf d}_k = (\gamma_w{\bf S}^H{\bf S}+\gamma_{2,k}{\bf I})^{-1}
\end{eqnarray}
\begin{eqnarray}
    \label{VAMPestimator_22}
     {\widetilde {\bf y}}_m = \gamma_w{\bf S}^H{\bf y}_m.
\end{eqnarray}
This completes the message passing from ${\bf h}_1$ to ${\bf h}_2$. Now, via Gaussian division, the message to be passed from the VN ${\bf h}_2$ to the FN $\delta$, $\mu_{{\bf h}_2\rightarrow\delta}$, can be similarly obtained as
\begin{eqnarray}
    \label{VAMPestimator_23}
     {\bf r}_{1,k+1} =  \frac{({\widehat{\bf h}}_{2,k}-\alpha_{2,k}{\bf r}_{2,k})}{1-\alpha_{2,k}}
\end{eqnarray}
\begin{eqnarray}
    \label{VAMPestimator_24}
     {\gamma_{1,k+1}} =\gamma_{2,k}{\frac{1-\alpha_{2,k}}{\alpha_{2,k}}}
\end{eqnarray}
%which are fu forwarded to the VN ${\bf h}_1$
where $k+1$ indicates the message is for the next $(k+1)$-th iteration. Convergence of the VAMP-based iterative channel estimation is determined from the normalized difference defined as $\xi_k=\frac{{\Vert {\bf r}_{1,k+1}-{\bf r}_{1,k}\Vert}_2^2}{{\Vert{\bf r}_{1,k+1}\Vert}_2^2}$. We stop the iteration when $\xi_k$ is lower than some predefined threshold $\xi_T$.
%A tolerance is empirically set as well. We stop the iteration the index is lower than the tolerance.

\subsection{\label{EM} Parameter Estimations using the EM Algorithm}

The parameter vector ${\bm \theta} = [{\bm \theta}_1,\gamma_w]^T$ involved in the VAMP based channel estimation is unknown and has to be estimated in practical use. In this section, we present a parameter estimation scheme based on the EM algorithm. The operation of the EM algorithm is very convenient in light of the use of VAMP for channel estimation, as the posterior probability required in the E step is already available in \eqref{VAMPestimator_4} and \eqref{VAMPestimator_14}. The EM algorithm is itself iterative and to
%can be used for obtaining expectation expression and further updating the ${\bm \theta}_1$ and $\gamma_w$ respectively\cite{Fletcher17}.
maintain a high efficiency, we make the iterations of the EM and VAMP coincide.

%The setting of hyper-parameters ${\bm \theta}$ is closely related to the performance of the VAMP.
%Due to the superiority of EM learning algorithm in estimating model hyper-parameters, it is chosen for VAMP based channel estimator hyper-parameters estimation.
%Different from the traditional form of an EM algorithm, the proposed scheme takes the self-iteration of the VAMP as the E-step to obtain information related to posterior probability %of ${\bf h}_m$ with given hyper-parameters and updates hyper-parameters as in a conventional M-step.
%Both \eqref{VAMPestimator_4} and \eqref{VAMPestimator_14} can be used as the posterior distribution for computing expectation function. \textcolor{blue}{We choose \eqref{VAMPestimator_4} to solve ${\bm \theta_1}$ and \eqref{VAMPestimator_14} to solve $\gamma_w$ in the EM method.} (\textcolor{red}{Why?})
%we first write expectation expression with respect to ${\bm \theta}_1$.
We next present the EM based parameter estimation in the $k$-th iteration. In the E-step, the expectation of $\text{ln}p(\uline{\bf h}_m,{\bf y}_m|{\bm \theta})$ is taken with respect to \eqref{VAMPestimator_4} \cite{Fletcher17}, leading to
\begin{eqnarray}
\begin{aligned}
    \label{VAMPestimator_25}
     &Q_1({\bm \theta}_1,{{\bm \theta}_{1,k-1}}) = \mathbb{E}[\text{ln}p(\uline{\bf h}_m,{\bf y}_m|{\bm \theta})|{\bf r}_{1,k},\gamma_{1,k},{\bm \theta}_{1,k-1}]\\
     & = \int{{\mathsf{ln}}(p({\uline{\bf h}}_m;{\bm \theta}_1)p({\bf y}_m|{\uline{\bf h}}_m;\gamma_w))b_1({\uline{\bf h}}_m|{\bf r}_{1,k},\gamma_{1,k},{\bm \theta}_{1,k-1})d{\uline{\bf h}}_m},
\end{aligned}
\end{eqnarray}
where ${{\bm \theta}_{1,k-1}}$ is the estimation of ${\bm \theta}_1$ in the $(k-1)$-th iteration. In the M-step, we seek an optimal parameter estimation, ${\bm \theta}_{1,k}$, that maximizes \eqref{VAMPestimator_25} as follows
\begin{eqnarray}
     \label{VAMPestimator_26}
     {\bm \theta}_{1,k} = \mathop{\arg\max}_{{\bm \theta}_1} \ \ Q_1({\bm \theta}_1,{{\bm \theta}_{1,k-1}})
\end{eqnarray}
The solution of ${\bm \theta}_{1,k}$ is obtained as follows
\begin{eqnarray}
     \label{VAMPestimator_32}
      \lambda_{k} = {\frac{1}{NL}}\sum_{i=1}^{NL}\pi_k(i)
\end{eqnarray}
\begin{eqnarray}
\begin{aligned}
     \label{VAMPestimator_37}
      \gamma_{h,k} %&= \frac{\sum_{i=1}^{NL}{\mathsf{lim}}_{\sigma \rightarrow 0}\int_{\vert {\uline h}_m(i)\vert \in {\overline{\mathcal B}}_{\sigma}}b_1({\uline h}_m(i)|r_{1,k}(i),\gamma_{1,k},{\bm \theta}_{1,k-1})d{\uline h}_m(i)}{\sum_{i=1}^{NL}{\mathsf{lim}}_{\sigma \rightarrow 0}\int_{\vert {\uline h}_m(i)\vert \in {\overline{\mathcal B}}_{\sigma}}{\vert {\uline h}_m(i)\vert}^2b_1({\uline h}_m(i)|r_{1,k}(i),\gamma_{1,k},{\bm \theta}_{1,k-1})d{\uline h}_m(i)}\\
      = \left({\frac{1}{\lambda_{k} NL}}\sum_{i=1}^{NL}\pi_k(i)({\vert\mu_k(i)\vert^2}+\nu_k)\right)^{-1}
\end{aligned}
\end{eqnarray}
where $\pi_k(i)$, $\mu_k(i)$, and $\nu_k$ are given by \eqref{VAMPestimator_6}, \eqref{VAMPestimator_7}, \eqref{VAMPestimator_8}, respectively.
The derivation is referred to Appendix B.

We continue to estimate $\gamma_w$ by constructing the following cost function in the E-step
\begin{eqnarray}
\begin{aligned}
    \label{VAMPestimator_38}
     &Q_2({\gamma_w,\gamma_{w,k-1}}) = \mathbb{E}[\mathsf{ln}(p({\uline{\bf h}}_m;{\bm \theta}_1)p({\bf y}_m|{\uline{\bf h}}_m;\gamma_w))|{\bf r}_{2,k},\gamma_{2,k},\gamma_{w,k-1}]\\
     & = \int{\mathsf{ln}(p({\uline{\bf h}}_m;{\bm \theta}_1)p({\bf y}_m|{\uline{\bf h}}_m;\gamma_w))b_2({\uline{\bf h}}_m|{\bf r}_{2,k},\gamma_{2,k},\gamma_{w,k-1})}d{\uline{\bf h}}_m
\end{aligned}
\end{eqnarray}
where $\gamma_{w,k-1}$ is the estimation of $\gamma_w$ in the $(k-1)$-th iteration.
In the M-step, we solve the following problem
\begin{eqnarray}
     \label{VAMPestimator_38_1}
     \gamma_{w,k} = \mathop{\arg\max}_{\gamma_w} \ \ Q_2({\gamma_w,\gamma_{w,k-1}}),
\end{eqnarray}
and the optimal estimation of $\gamma_w$ is obtained as
\begin{align}
     \gamma_{w,k}= \bigg(\frac{1}{K_p}{\Vert {\bf y}_m - {\bf W}{\widehat{\bf h}}_{2,k} \Vert}^{2}+\frac{1}{K_p}{\gamma^{-1}_{w,k-1}}\sum_{n=1}^{R}(\frac{{\vert s_n \vert}^{2}}{{\vert s_n \vert}^{2}+{\gamma^{-1}_{w,k-1}}{\vert s_n \vert}^{2}})\bigg)^{-1}\label{VAMPestimator_42}
\end{align}
where ${\widehat{\bf h}}_{2,k}$ is given in \eqref{VAMPestimator_19}. Detailed derivation procedure is found in Appendix C.

Last, initialization of the three parameters $\gamma_{w}$, $\lambda$, and $\gamma_{h}$, is necessary for the operation of the EM based iterative parameter estimation. The initial value of $\gamma_{w}$ is set as\cite{Vila13}
\begin{eqnarray}
     \label{VAMPestimator_43}
     \gamma_{w,0} = \frac{(1+\zeta)K_p}{{\Vert {\bf y}_m \Vert}_{F}^{2}}
\end{eqnarray}
where $\zeta$ as the signal-to-noise ratio (SNR) can be estimated from observation data or is just heuristically set. A heuristic initial value for the normalized sparsity, $\lambda$, has been suggested as $\lambda_0=0.95$. %empirically
%\begin{eqnarray}
%     \label{VAMPestimator_44}
%     \lambda_0 = \frac{K_p}{NL}\max\limits_{b>0}\frac{1-\frac{2NL}{K_p}[(1+b^2)G(-b)-g(b)]}{1+b^2-2[(1+b^2)G(-b)-g(b)]}
%\end{eqnarray}
%where $G(\cdot)$ and $g(\cdot)$ denote the cumulative distribution function and the probability density function of the standard Gaussian distribution, respectively.
The initial value for $\gamma_{h}$ can be set as \cite{Vila13}
\begin{eqnarray}
     \label{VAMPestimator_45}
     \gamma_{h,0} = \frac{{\Vert {\bf W} \Vert}_{F}^{2}\lambda_0}{{\Vert {\bf y}_m \Vert}_{F}^{2}-K_p{{\gamma_{w,0}}^{-1}}}
\end{eqnarray}
The procedure of EM-VAMP channel estimation is finally summarized in Algorithm 1.
\IncMargin{1em} % 使得行号不向外突出
\begin{algorithm}[htbp]
\normalem
\begin{spacing}{1.5}

    \SetAlgoNoLine % 不要算法中的竖线
    \vspace{2mm}
    \SetKwInOut{Input}{\textbf{Require}}\SetKwInOut{Output}{\textbf{Return}} % 替换关键词

   % \Input{Matrix ${\bf S}_p$; measurements ${\bf Y}_p$; estimated noise precision $\gamma_w$; and
    % number of iterations $K_{it}.$\\}
    %\BlankLine
    {\bf Input}: The measurement matrix in form of SVD decomposition ${\bf W}={\bf U}{\bf S}{\bf V}^{H}$, and %, {R} = rank({\bf W})
    the observation vector ${\bf y}_m$. \\ %${\widetilde {\bf y}}_m = \gamma_w{\bf S}^H{\bf y}_m$ as in \eqref{VAMPestimator_22}.\\
    %The observation vector ${\bf y}_m$, the measurement matrix ${\bf W}$;
    %estimated noise precision $\gamma_w$;\\
    {\bf Initialization}: Set ${\bf r}_{1,1}={\bf 0}$, $\gamma_{1,1}=1$, $\lambda_0=0.95$, and $\gamma_{w,0}$ and $\gamma_{h,0}$ according to \eqref{VAMPestimator_43} and \eqref{VAMPestimator_45}. Set the maximum number of iterations $K_{\text{max}}$ and stopping threshold $\xi_T$;\\
      \For {$k = 1, 2, \cdots, K_{\text{max}}$}{
        \ $\pi_k(i) = \left(1+\frac{(1-\lambda_{k-1})\mathcal {CN}({r_{1,k}(i)};0,{\gamma_{1,k}}^{-1})}{\lambda_{k-1}\mathcal {CN}({r_{1,k}(i)};0,({\gamma_{1,k}}^{-1}+\gamma_{h,{k-1}}^{-1}))}\right)^{-1}$\\
\   ${\widehat{\bf h}}_{1,k} = {\bm \pi}_k\odot\frac{\gamma_{1,k} {\bf r}_{1,k}}{\gamma_{1,k}+\gamma_{h,{k-1}}},$ %=  {\bm \pi}_k{\bm \mu}_k
\   $ {\alpha_{1,k}} = \left\langle{\bm \pi_k}\frac{\gamma_{1,k}}{\gamma_{1,k}+\gamma_{h,k-1}}\right\rangle,$ $\nu_k = \frac{1}{\gamma_{1,k}+\gamma_{h,k-1}}$\\
\   $ {\bf r}_{2,k} =  \frac{({\bf{\widehat h}}_{1,k}-\alpha_{1,k}{\bf r}_{1,k})}{1-\alpha_{1,k}},$
\   $ \gamma_{2,k} =  \gamma_{1,k}{\frac{1-\alpha_{1,k}}{\alpha_{1,k}}}$\\
\   $\lambda_{k} = {\frac{1}{NL}}\sum_{i=1}^{NL}{\pi_{k}}(i), \gamma_{h,k}=\big({\frac{1}{\lambda_{k} NL}}\sum_{i=1}^{NL}{\pi_{k}(i)({\vert\mu_{k}(i)\vert^2}+\nu_{k})}\big)^{-1}$\\

\   ${\bf d}_k = (\gamma_{w,k-1}{\bf S}^H{\bf S}+\gamma_{2,k}{\bf I})^{-1}$, ${\widetilde {\bf y}}_m = \gamma_{w,k-1}{\bf S}^H{\bf y}_m$\\
\   ${\widehat{\bf h}}_{2,k} = {\bf V}{\bf d}_k({\widetilde {\bf y}}_m + \gamma_{2,k}{\bf V}^H{\bf r}_{2,k}),$
\   $\alpha_{2,k} = \frac{1}{R}\sum_{n}\frac{\gamma_{2,k}}{\gamma_{w,k-1}{s}_n^2+\gamma_{2,k}}~\text{with}~{s}_n = [{\bf S}]_{nn}$\\
\   ${\bf r}_{1,k+1} =  \frac{({\widehat{\bf h}}_{2,k}-\alpha_{2,k}{\bf r}_{2,k})}{1-\alpha_{2,k}},$
\   ${\gamma_{1,k+1}} =\gamma_{2,k}{\frac{1-\alpha_{2,k}}{\alpha_{2,k}}}$\\
\   $ \gamma_{w,k} = \bigg(\frac{1}{K_p}{\Vert {\bf y}_m - {\bf W}{\widehat{\bf h}}_{2,k} \Vert}^{2}+\frac{1}{K_p}{\gamma^{-1}_{w,k-1}}\sum_{n=1}^{R}(\frac{{\vert s_n \vert}^{2}}{{\vert s_n \vert}^{2}+{\gamma^{-1}_{w,k-1}}{\vert s_n \vert}^{2}})\bigg)^{-1}$\\
\    $\textbf{if}\quad{\Vert {\bf r}_{1,k+1}-{\bf r}_{1,k}\Vert}_2^2 \leq \xi_T{\Vert{\bf r}_{1,k+1}\Vert}_2^2$\\
\       $\qquad \text{break}.$\\
\   $\textbf{end if}$\\
    }
    {\bf Output}: ${\bf{\widehat h}}_{1,K_{f}}$ where $K_f$ is the actual number of iterations executed. %in $K_{EM}$-th EM iteration.
    \caption{EM-VAMP Channel Estimation}
\end{spacing}
\end{algorithm}
\DecMargin{1em}

\subsection{\label{Complexity} Complexity Analysis and Comparison}
%\textcolor{red}{ It should be noted that EM-VAMP and SBL both use the assumed prior and likelihood probability to obtain the posterior probability, and then update the hyperparameters in the posterior probability iteratively to obtain the nearly accurate posterior probability. Finally the posterior mean is chosen as the estimation. The diversity lies in that EM-VAMP adopts the thought of message passing and uses the posterior probability calculated in the process of message passing to update the hyperparameters while SBL directly obtains the posterior probability based on the prior probability and likelihood to update the hyperparameters, which leads to the difference in performance.}
In this subsection, we analyze the computational complexity of the proposed EM-VAMP-CE scheme and compare it with those of existing schemes, including the LS, LMMSE, OMP, and SBL CEs.
It is noted existing SBL CE \cite{Qiao18} adopts a prior probability model in which channel taps follow independent Gaussian distributions with different variance. Such a model is more accurate, but involves a large number (proportional to the channel length) of hyperparameters.
%SBL-simplify, VAMP, EM-VAMP CESs is made.
%\textcolor{red}{For the fairness of comparison, we also test the SBL-CE with Bernoulli-Gaussian prior as that for the VAMP, and called the algorithm as SBL-BG-CE.} %In this way, complexity can be heavily saved.
The complexity is measured in the number of complex multiplications (CMs). For convenience, a division operation is treated as a multiplication operation, and all the addition operations are excluded for their low computation cost\cite{Tao18}. As the complexity of most schemes is related to the choice of pilot sequences, both cases of orthogonal and non-orthogonal pilot sequences are investigated. In the case of non-orthogonal pilot sequences, the SVD ${\bf W}= {\bf U}{\bf S}{\bf V}^{H}$ is used for LS, LMMSE, EM-VAMP CEs to reduce complexity and the complexity of the SVD is not included. For the EM-VAMP-CE, the complexity of EM-based parameter estimation is much lower than the VAMP, thus is not included.
%Apart from the types of algorithms, the design of pilot sequence also has an important influence on the computational complexity of channel estimation.
%Here, we consider both equipowered, equispaced and equipowered, equispaced, phase shift orthogonal pilots design\cite{Barhumi03}.
%Use of non-orthogonal and leads to different complexity. The design of orthogonal pilot sequences across transducers is referred to \cite{Barhumi03}

The complexity comparisons are listed in Table~\ref{complexity comparsion}, where $K_h$, $K_f$ and $K_s$ are the sparse level of ${\uline{\bf h}}_m$, the iteration number of the EM-VAMP-CE, and the number of iterations for the SBL-CE, respectively. Clearly, use of orthogonal pilot sequences leads to complexity saving for all but the OMP CEs.
In the case of non-orthogonal pilot sequences, the EM-VAMP-CE has a much lower complexity than the SBL-CE.
%In the case of orthogonal pilot sequences, the SBL-CE and EM-VAMP-CE are comparable in complexity. %The OMP-CE enjoys the lowest complexity in both orthogonal and non-orthogonal pilot sequences, attributed to its direct use of the channel sparsity level.
%Since the transmitted symbols are known, the $\bf W$ is known. The SVD decomposition of ${\bf W}= {\bf U}{\bf S}{\bf V}^{H}$ can be performed once in advance in the practical engineering applications. Therefore this operation is not considered in the computational complexity. The simplified forms of LS-CE and LMMSE-CE after SVD decomposition ${\bf W}={\bf U}{\bf S}{\bf V}^{H}$ are as follows.
%\begin{eqnarray}
%\begin{aligned}
%     \label{LS LMMSE SVD}
%     &\uline{\widehat{\bf h}}_{m,LS} = {\bf V}({\bf S}^{H}{\bf S})^{-1}{\bf S}^H{\bf U}^H{\bf y}_m\\
%     &\uline{\widehat{\bf h}}_{m,LMMSE} = {\bf V}{\bf S}^{H}({\bf S}{\bf S}^{H}+\frac{\sigma^2_w}{\sigma^2_h}{\bf I})^{-1}{\bf U}^H{\bf y}_m
%\end{aligned}
%\end{eqnarray}
%where $\sigma^2_w$ is noise variance and $\sigma^2_h$ is average power of channel taps (the average power of all taps in the channel impulse response is assumed to be consistent).
\begin{table}[!ht] % 表格开始
\renewcommand\arraystretch{1.5}
\centering % 表格居中显示
\caption{Complexity comparison (Number of CMs)} % 表头标题
\label{complexity comparsion} % 表格标签，便于引用

\begin{tabular}{ccc} % c表示单元格内容居中，l表示靠左，这里有两列，所以是cc，如果是三列就是ccc或cll，根据自己

\toprule[1.5pt] % 顶部的线，这里可以定义粗细、toprule{1.5ptx}
\multicolumn{1}{m{5cm}}{\centering \bf{CE methods}}& % 中间的1.5cm表示第一列宽度
\multicolumn{1}{m{5cm}}{\centering \bf{Non-orthogonal pilot sequences}}& %Equipowered, Equispaced pilots
\multicolumn{1}{m{5cm}}{\centering \bf{Orthogonal pilot sequences}}\\

\midrule % 中间的线
%LS-CE&$\mathcal{O}(MN^2_p+3MNL+MN^2L^2)$\\ % &用来区分列，\\用来区分行
LS&$\mathcal{O}(MN^2_p+MN^2L^2)$&$\mathcal{O}(MNK_pL)$\\
%LMMSE-CE&$\mathcal{O}(MN^2_p+2MK_p+MN^2L^2+2MNL)$\\
LMMSE&$\mathcal{O}(MN^2_p+MN^2L^2)$&$\mathcal{O}(MNK_pL)$\\
OMP&$\mathcal{O}(MN^2_pK_h+\frac{MK^3_h}{3})$&$\mathcal{O}(MN^2_pK_h+\frac{MK^3_h}{3})$\\
SBL&$\mathcal{O}(\frac{MN^3K_sL^3}{2}+2MN^2K_sL^2K_p)$&$\mathcal{O}(MNK_sK_pL)$\\
%SBL-BG-CE&$\mathcal{O}(MK_sN^2_p+MNK_sK_pL)$&$\mathcal{O}(MNK_sK_pL)$\\
%VAMP-CE&$\mathcal{O}(MN^2L^2K_f)$&$\mathcal{O}(MK_pNLK_f)$\\
EM-VAMP&$\mathcal{O}(MN^2K_fL^2)$&$\mathcal{O}(MNK_f K_pL)$\\

\bottomrule[1.5pt]
\end{tabular}
\end{table}

\section{Simulation results\label{SimulationResult}}

\begin{figure}[!ht] %[!thbp]
   \hspace{-0.4cm}\centering
   \centering
   \includegraphics[width=13.5cm]
   {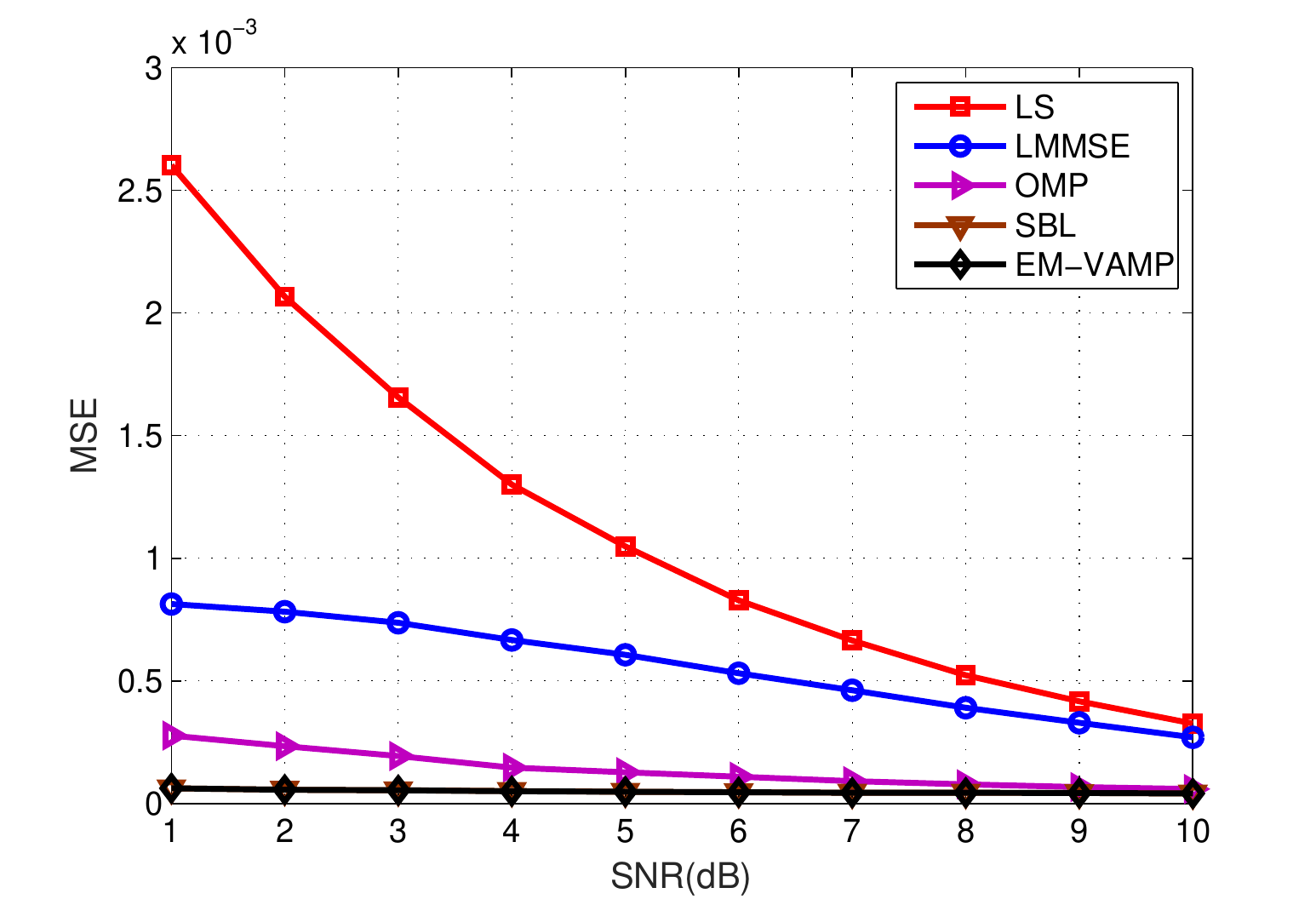}
   \caption{Comparison of channel estimation MSE at different SNRs.}
   \label{SimChannelMSE}
\end{figure}
In this section, we study the performance of the proposed EM-VAMP-CE through simulations. A $2\times 6$ MIMO OFDM UWA communication system is considered.
The twelve subchannels are taken from the SPACE08 experiment \cite{Tao10}, and have the same length of $L=100$. %\textcolor{red}{sparse level of the channel?}
%A rate-$\frac{1}{2}$ convolutional channel code was adopted. Each transducer transmits a total of 80 blocks.
Each OFDM block contains $K=1024$ subcarriers including $K_d=768$ data subcarriers and $K_p=256$ equally-spaced pilot subcarriers for channel estimation. The pilot sequences across the two transducers are orthogonal to each other \cite{Barhumi03}. %The data and pilot symbols are modulated by QPSK constellations.
We evaluate the performance of LS, LMMSE, OMP, SBL, and EM-VAMP CEs in terms of the channel estimation mean square error (MSE).
%For the OMP, what about the termination condition? For the SBL, what about the initialization? For the LMMSE, what about the channel covariance matrix? }

 % in the hydrophone SNR range of 1-8 dB
The MSE comparison is shown in Fig.~\ref{SimChannelMSE}. From the figure, the EM-VAMP-CE and SBL-CE achieve similar channel estimation accuracy that is higher than the LS, LMMSE, and OMP CEs.

%At the same time, although the Bayesian estimation accuracy of EM-VAMP and SBL for channel is similar in simulation, the prior model of EM-VAMP has fewer hyperparameters, which is more concise and easy to adjust.

\section{Experimental results\label{ExperimentResult}}

The proposed EM-VAMP-CE has also been tested by experimental data collected in two independent UWA communication experiments: XM16 and SPACE08. The results are presented in Section~\ref{Xiamen} and Section~\ref{Space08}, respectively.

\subsection{\label{Xiamen} The XM16 Experiment}

%The QPSK modulation was used.
This experiment was conducted at the Wuyuan Bay, Xiamen, China, in Dec. 2016. It was a multichannel communication with four hydrophone elements on the receiver side. The carrier frequency was 16 kHz and the bandwidth was 4 kHz. A rate-$\frac{1}{2}$ LDPC channel code was adopted. Every transmission packet consisted of 6 OFDM blocks each having
$K=1280$ subcarriers, among which $K_d=1024$ were data subcarriers, $K_p=160$ were equally-spaced pilot subcarriers, and 96 were null subcarriers. The channel length was measured as $L=150$. %\textcolor{red}{The tolerance $\xi_T$ is set to $1\times10^{-3}$}. %, and the signal to noise ratio (SNR) on each hydrophone is about 20 dB

\begin{table}[!ht]
\renewcommand{\arraystretch}{1.5}
\begin{center}
\caption{Three sets of parameter initializations.}
\label{hyperparameter configuration}
%{\begin{tabular}{|c||p{0.5in}|p{0.5in}|p{0.5in}|}\hline
{\begin{tabular}{|c||p{1.5cm}<{\centering}|p{1.5cm}<{\centering}|p{1.5cm}<{\centering}|p{1.5cm}<{\centering}|p{1.5cm}<{\centering}|p{1.5cm}<{\centering}|p{2cm}<{\centering}|}\hline
\diagbox{\bf Hyperparameter}{\bf Set Index} & 1 & 2 & 3
%\backslashbox
%\\\hline\hline ${\lambda}$ & 0.9 & 0.7 & 0.5
%\\\hline ${\gamma}_h$ & 253.8 & 157.5 & 84.8
%\\\hline ${\gamma}_w$ & 79.9 & 41.7 & 21.4 \\\hline
\\\hline\hline ${\lambda}$ & 0.95 & 0.75 & 0.55
\\\hline ${\gamma}_h$ & 298.7 & 200 & 100
\\\hline ${\gamma}_w$ & 86.4 & 60 & 20 \\\hline
\end{tabular}}{}
\end{center}
\end{table}
\begin{figure}[!ht] %[!thbp]
\centering
 % \subfigure[]{
%    \includegraphics[scale=0.6]{lamdaIteration.eps}
%    \hspace{-0.6cm}%in表示英寸，\hspace 调整图片水平距离
%    \includegraphics[scale=0.6]{gammahIteration.eps}
%    \hspace{-0.6cm}
%    \includegraphics[scale=0.6]{gammawIteration.eps}
    \includegraphics[scale=0.7]{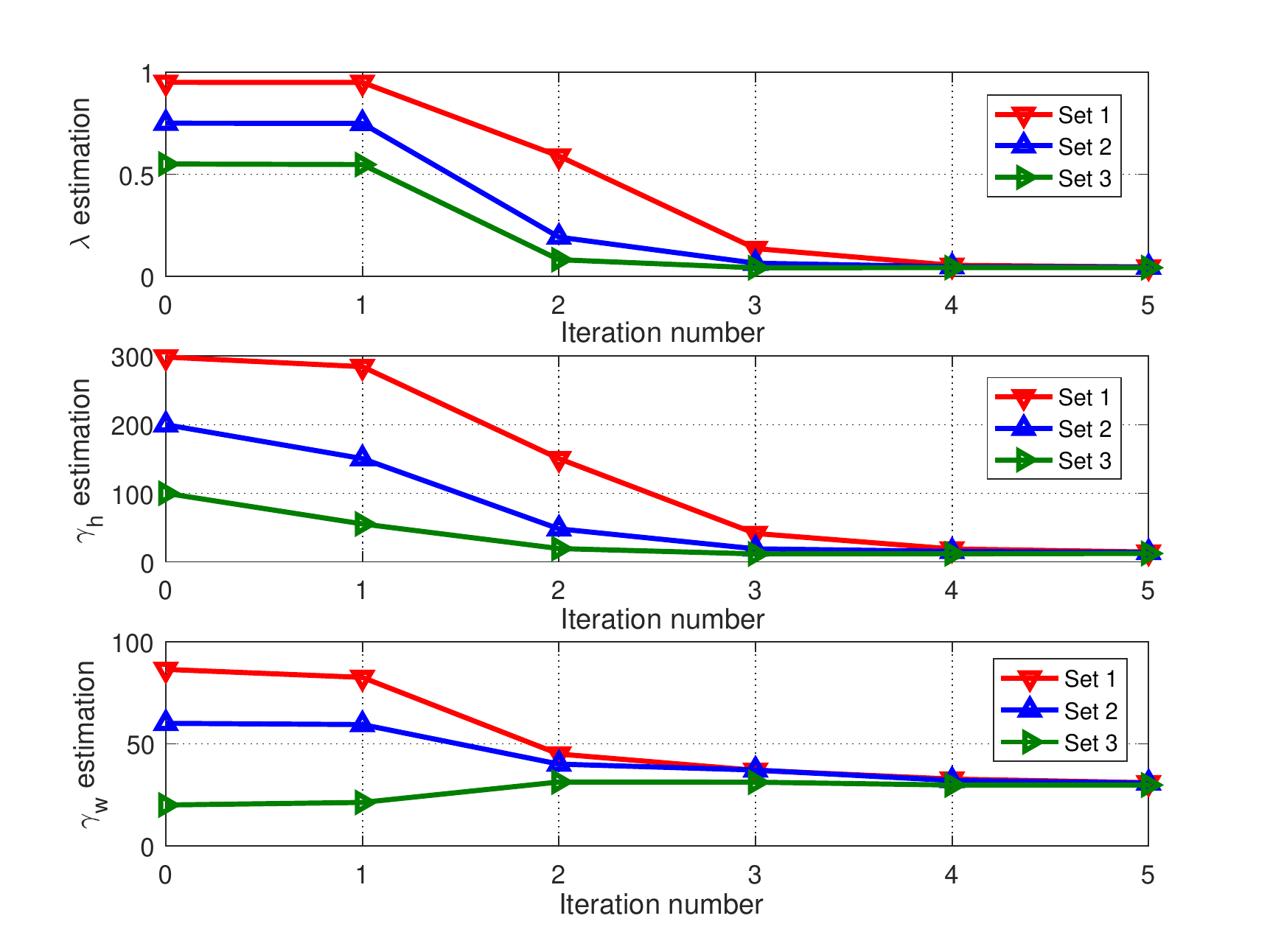}
  \caption{Evolution of the parameter estimations.}
  \label{hyperparameters convergence}
\end{figure}
%It is obvious that hyperparameters will eventually converge to the same values over the iterations of EM-VAMP-CE if we set different initial values.
%Adaptive estimation of the hyperparameters with EM-VAMP-CE iteration is first investigated.shows three different initial value configuration for. shows the convergence of all the hyperparameters.

\begin{figure}[!ht]
\centering
%\subfigure[Block 1]{
%\includegraphics[width=7cm]{XMchannelPlotBlock1.eps}
%%\caption{Block 1}
%}
%\quad
%\subfigure[Block 2]{
%\includegraphics[width=7cm]{XMchannelPlotBlock2.eps}
%}
%\quad
%\subfigure[Block 3]{
%\includegraphics[width=7cm]{XMchannelPlotBlock3.eps}
%}
%\quad
%\subfigure[Block 4]{
%\includegraphics[width=7cm]{XMchannelPlotBlock4.eps}
%}
   \includegraphics[width=15.5cm]{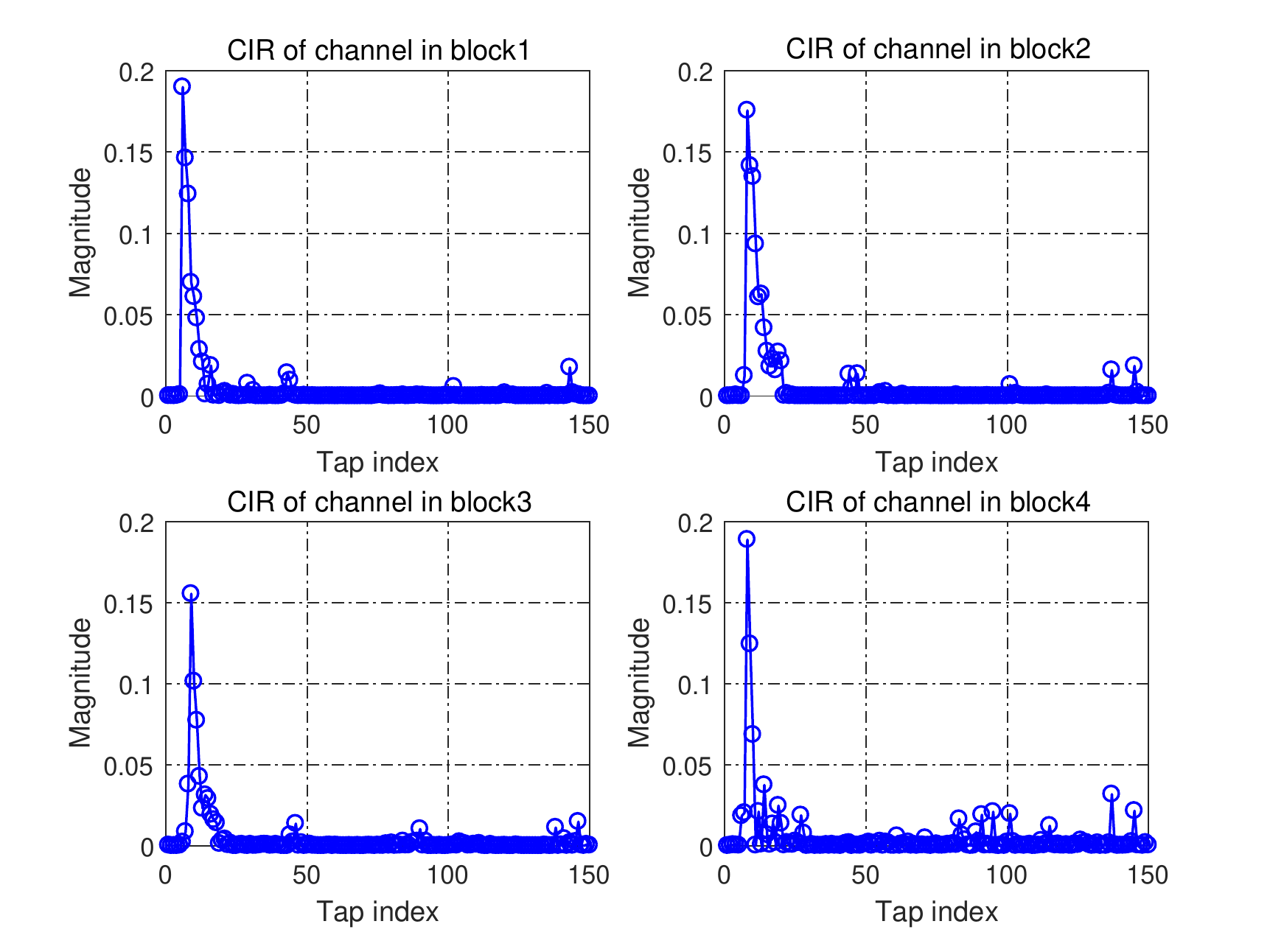}
\caption{Estimated channel impulse responses via EM-VAMP-CE for four blocks (XM16 experiment).}
\label{XMchannelPlot}
\end{figure}
First, we demonstrate the learning of the parameters: ${\lambda}$, ${\gamma}_h$, and ${\gamma}_w$.
%First, we demonstrate the EM-based hyperparameter estimation for the first block.
%As there is no prior knowledge available, the three parameters were initialized empirically.
Three sets of initial values, as listed in Table~\ref{hyperparameter configuration}, were adopted to study the impact of initialization on the EM-based parameter estimation. The first set was obtained according to the suggested initialization introduced at the end of Section.~\ref{EM}. Evolutions of the parameter estimations over EM iterations are shown in Fig.~\ref{hyperparameters convergence}. Clearly, all three parameter estimations converge to almost identical values in four iterations, despite their initial settings. The insensitivity to initialization is desirable for practical use. The converged sparsity rate $\lambda$ is about 0.05, indicating the underlying channel is very sparse. This is corroborated by the estimated channels of four OFDM blocks on the first receive hydrophone, demonstrated in Fig.~\ref{XMchannelPlot}.

\begin{figure}[!ht] %[!thbp]
  \centering
  \subfigure[MSE of estimated symbols comparison at different SNRs.]{\includegraphics[width=5in]{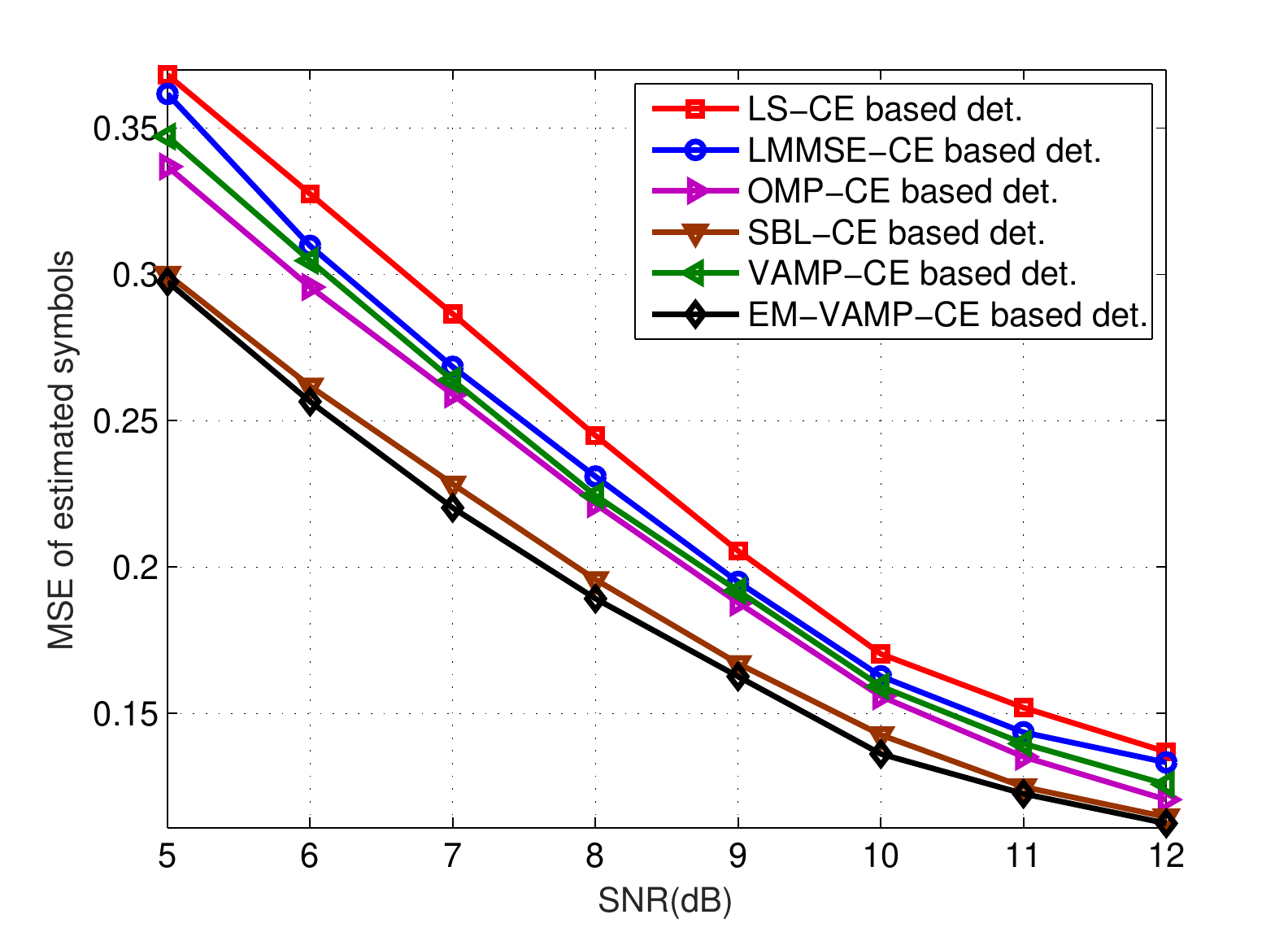}}
  \subfigure[Raw BER comparison at different SNRs.]{\includegraphics[width=5in]{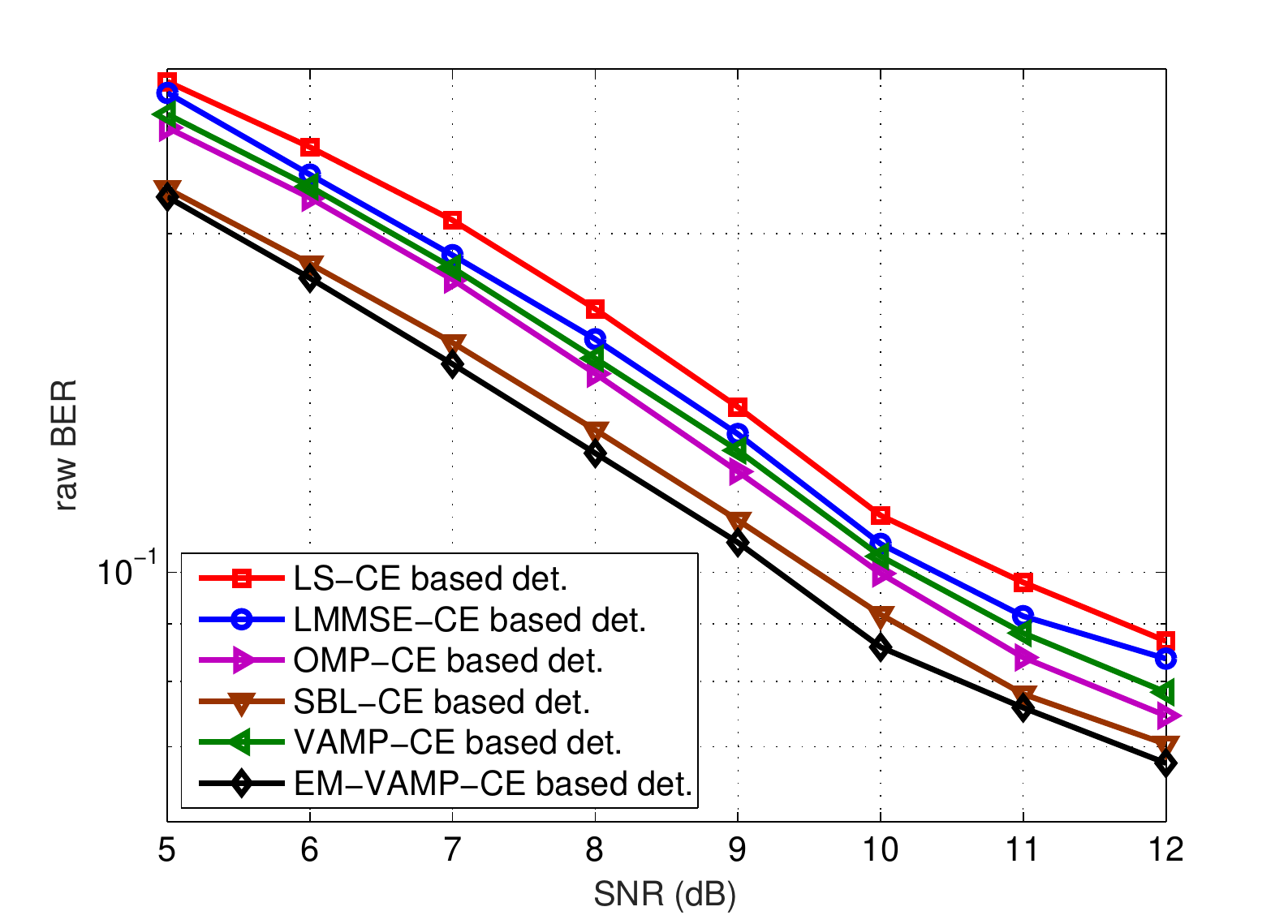}}
  \caption{MSE of estimated symbols and raw BER comparisons (XM16 experiment).}
  \label{performanceComparison}
\end{figure}
\begin{figure}[!ht] %[!thbp]
   \hspace{-0.4cm}\centering
   \centering
   \includegraphics[width=13.5cm]
   {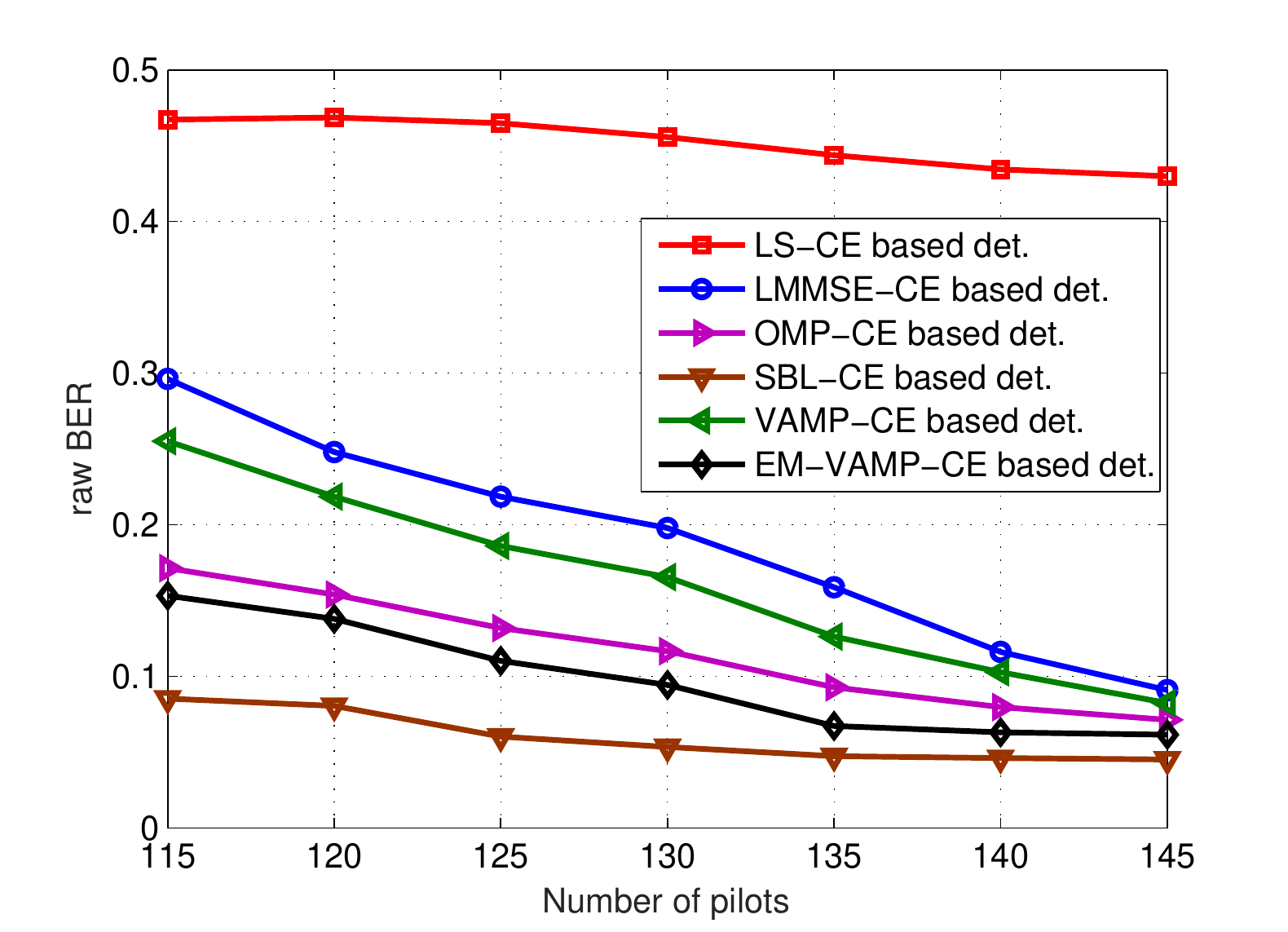}
   \caption{Raw BER comparison with different number of pilot symbols (XM16 experiment).}
   \label{XMPilotsNumber}
\end{figure}
Next, performance of the proposed EM-VAMP-CE is presented and compared with estimations based on LS, LMMSE, OMP \cite{Tropp07}, and SBL\cite{Wipf04} algorithms. The VAMP CE with predetermined hyperparameters (which are set according to the discussion at the end of Section.~\ref{EM}) was also investigated. The empirical settings are also used to initialize the EM-VAMP-CE. As the real channel is unknown, the MSE of estimated symbols and raw bit error rate (BER), that is the bit error rate before decoding, were adopted as the performance metrics.
%we adopt block average bit error rate before decoding (called raw BER hereafter) and the MSE of symbol estimation as performance criteria. As the true channel is unknown, indirect performance metric of raw bit error rate (BER), that is BER before decoding, was adopted.
The SNR of the originally received data is about 20 dB. To enable an investigation under different SNR levels, artificial Gaussian noise was added to achieve an SNR range of 5-12 dB. %The number of pilot symbols used is 160.
The performance comparison was shown in Fig.~\ref{performanceComparison}, where the EM-VAMP-CE outperforms all other CEs.
%indirect criteria to evaluate the CE performance. Specifically,. LS, LMMSE, OMP\cite{Tropp07}, and VAMP\cite{Rangan19} CEs are used to compared with EM-VAMP-CE, where VAMP-CE represents VAMP with empirical hyperparameters based channel estimation.
%As shown in the Fig.~\ref{comparisonSNR1}, the performance of EM-VAMP-CE outperforms LS, LMMSE, OMP, and VAMP CEs.

%We want to see how the reduction in the number of pilots affects each CE algorithm.
Last, we investigate the impact of pilot overhead on the channel estimation performance. According to \cite{Tao18}, the minimum number of pilots required by the LS channel estimation is equal to the channel length $L=150$, so as to avoid an under-determined problem. Attributed to the channel sparsity, however, sparsity-aware algorithms work decently even with less pilots. Therefore, we focused on the case where the number of pilots is less than 150. The results are shown in Fig.~\ref{XMPilotsNumber}, where the SNR is 20dB and the raw BER is used as the performance metric. From the figure, all but the LS CEs work properly. The EM-VAMP-CE consistently outperforms all estimation methods except for the SBL-CE, when the number of pilots goes from 115 to 145.
%After all, the AMP algorithm was initially proposed as a reconstruction method for solving compressive sensing problem.
The LS channel estimation completely failed, due to insufficient pilots. %make the channel estimation problem under-determined.
%The number of pilots changed from 120 to 155. Raw BER is chosen as our indicator.. shows that the consumes much less pilot resources to achieve the same performance.

\subsection{\label{Space08} The SPACE08 Experiment} %Experience II: SPACE08

The second experiment was performed near Martha's Vineyard, Edgartown, MA, in 2008 \cite{Tao10}. The carrier frequency was 13 kHz and the occupied bandwidth was 9.7656 kHz. The QPSK modulation and a rate-$\frac{1}{2}$ convolutional channel code were adopted. The transmit equipment consisted of four transducers and the receive equipment consisted of twelve hydrophones. Experimental results for a $2\times 6$ MIMO transmission are discussed in the following. Each transmission packet consisted of 8 QPSK blocks per transducer. Each block carried $K = 1024$ symbols with $K_p = 256$ being pilots. Non-orthogonal pilot sequences were adopted across the two transducers. In total, four received packets were processed. The lengths of all subchannels were measured within 100, and the SNR on each hydrophone was about 10 dB. %The tolerance $\xi_T$ is again set to $1\times10^{-3}$.  and the symbol period was about 0.1 ms

%\begin{table}[!ht]
%\renewcommand{\arraystretch}{0.6}
%\begin{center}
%\caption{\bf Hyperparameters estimation for different iteration}
%\label{hyperparameter estimation}
%%{\begin{tabular}{|c||p{0.5in}|p{0.5in}|p{0.5in}|}\hline
%{\begin{tabular}{|c||p{1.5cm}<{\centering}|p{1.5cm}<{\centering}|p{1.5cm}<{\centering}|p{1.5cm}<{\centering}|p{1.5cm}<{\centering}|p{1.5cm}<{\centering}|p{2cm}<{\centering}|}\hline
%\diagbox{\bf Hyperparameter}{\bf EM-VAMP-CE\\\bf iteration} & 1 & 2 & 3 & 4 & 5
%%\backslashbox
%\\\hline\hline ${\bm \lambda}$ & 0.998 & 0.997 & 0.993 & 0.982 & 0.959
%\\\hline ${\bf \gamma_h}(10^4)$ & 6.460 & 3.335 & 3.394 & 3.452 & 3.400
%\\\hline ${\bf \gamma_w}(10^3)$ & 1.114 & 0.768 & 0.779 & 0.800 & 0.803\\\hline
%\end{tabular}}{}
%\end{center}
%\end{table}

\begin{figure}[!ht] %[!thbp]
   \hspace{-0.4cm}\centering
   \centering
   \includegraphics[width=13.5cm]
   {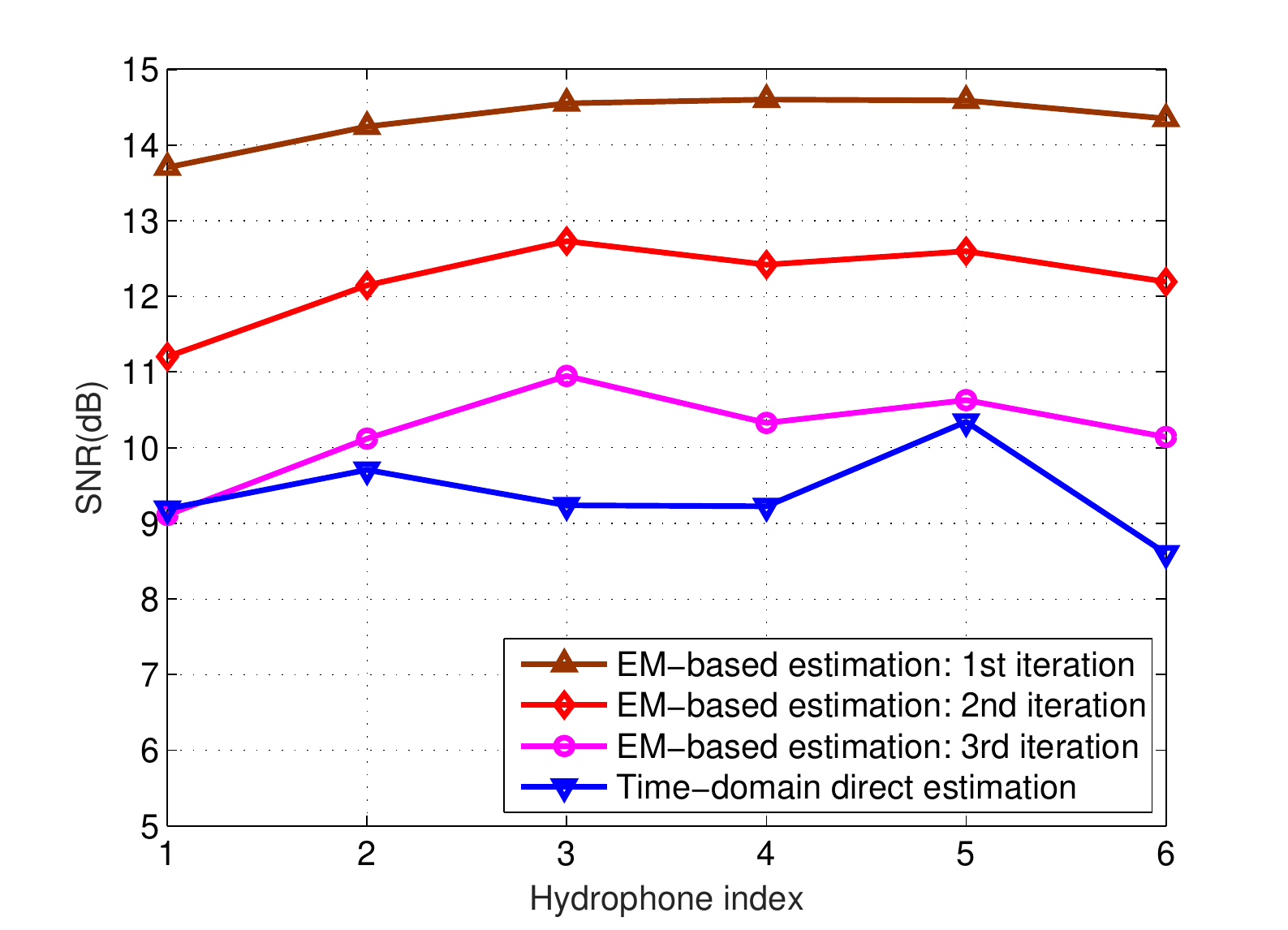}
   \caption{A comparison of the SNR estimations (SPACE08 experiment).}
   \label{SNR estimation}
\end{figure}

We start with investigating the effectiveness of the EM-based parameter estimation. We focus on the estimation of $\gamma_w$ or equivalently, the SNR on the receiver side. The reason is that the SNR can be directly estimated with received time-domain signal, attributed to the gaps inserted among OFDM blocks.
%by computing the power ration between the signal and noise part  of the received time domain sequence.
The time-domain direct estimation can be used as a reference. In Fig.~\ref{SNR estimation}, the SNR estimations on six receive hydrophones were compared. One can see that at convergence, the estimation via the EM algorithm is close to that estimated from the time series directly. %shows the results of the SNR estimation by above methods respectively for different receivers.
%Using the estimated subchannel impulse response and noise precision obtained by EM-VAMP-CE, we can approximately get SNR at each receiver.
%\begin{figure}[!ht] %[!thbp]
%  \centering
%  \subfigure[SNR estimation for different blocks]{\includegraphics[width=4in]{figure21.eps}}
%  \subfigure[SNR estimation for different receivers]{\includegraphics[width=4in]{figure22.eps}}
%  \caption{SNR estimation for different methods }
%  \label{SNR estimation}
%\end{figure}

\begin{figure}[!ht]
\centering
%\subfigure[Block 1]{
%\includegraphics[width=7cm]{Sp08ChannelPlotT1H1B1.eps}
%%\caption{fig1}
%}
%\quad
%\subfigure[Block 1]{
%\includegraphics[width=7cm]{Sp08ChannelPlotT2H1B1.eps}
%}
%\quad
%\subfigure[Block 7]{
%\includegraphics[width=7cm]{Sp08ChannelPlotT1H1B7.eps}
%}
%\quad
%\subfigure[Block 7]{
%\includegraphics[width=7cm]{Sp08ChannelPlotT2H1B7.eps}
%}
   \includegraphics[width=15.5cm]{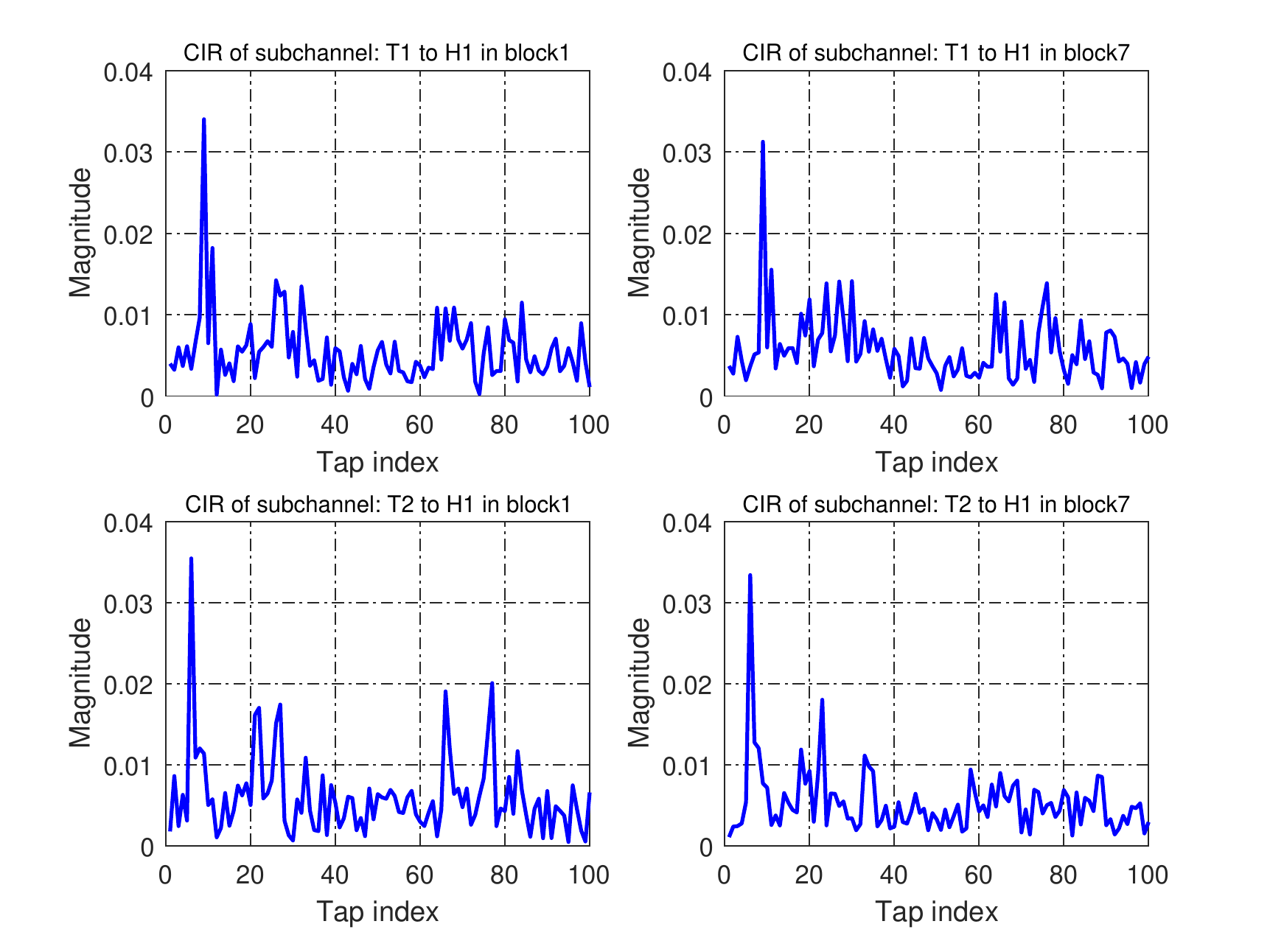}
\caption{The estimated channel impulse responses via EM-VAMP-CE (SPACE08 experiment).}
\label{Sp08ChannelPlot}
\end{figure}
The sparsity rate estimation for this experiment was about 0.2, indicating the underlying channel is less sparse compared with that of the previous XM16 experiment. In Fig.~\ref{Sp08ChannelPlot}, examples of estimated subchannels between transducers 1,2 (denoted by T1 and T2) and hydrophone 1 (denoted as H1) for blocks 1 and 7, are shown.

\begin{figure}[!ht] %[!thbp]
  \centering
  \subfigure[The symbol MSE comparison at different SNRs.]{\includegraphics[width=5in]{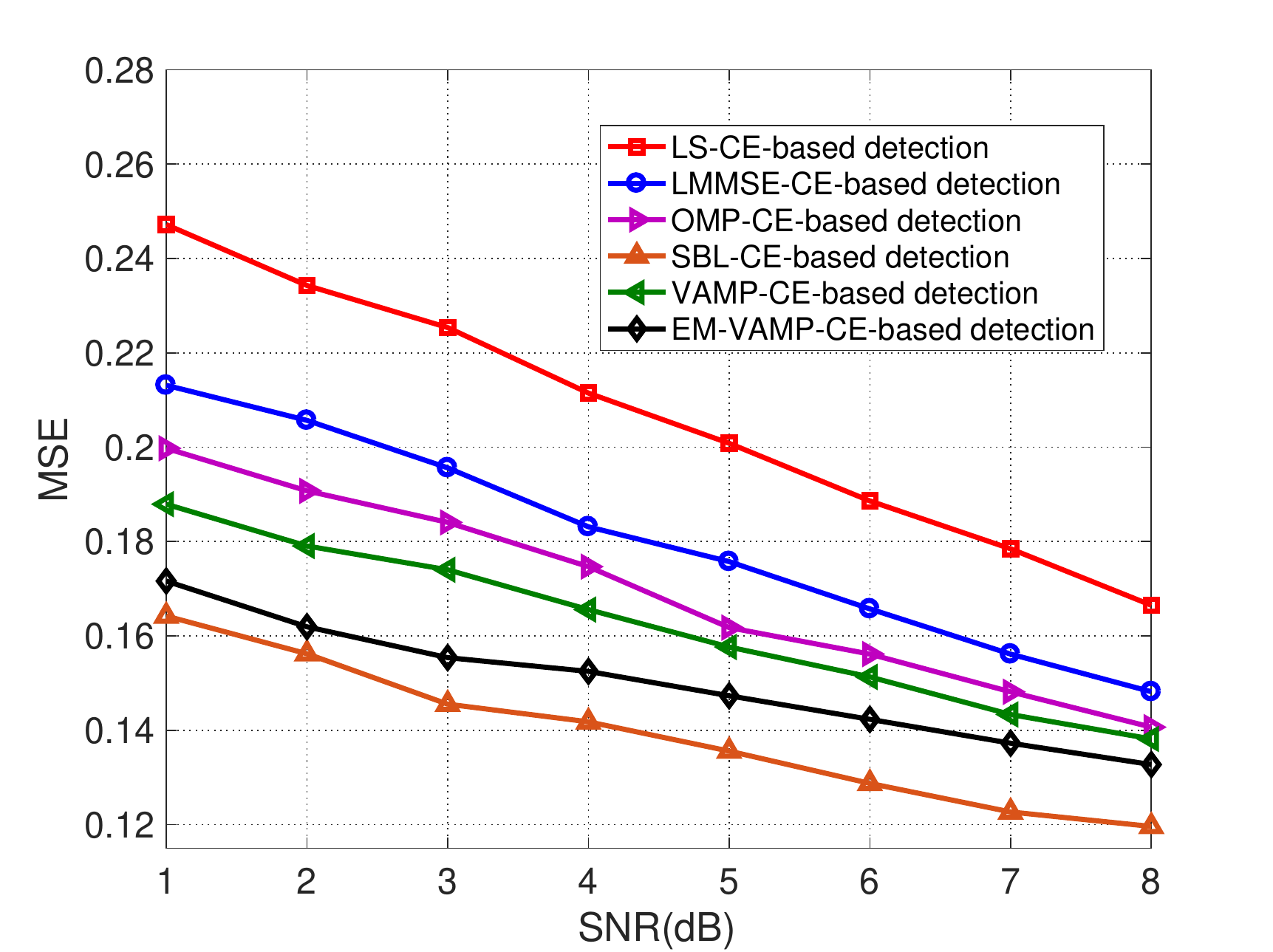}}
  \subfigure[The raw BER comparison at different SNRs.]{\includegraphics[width=5in]{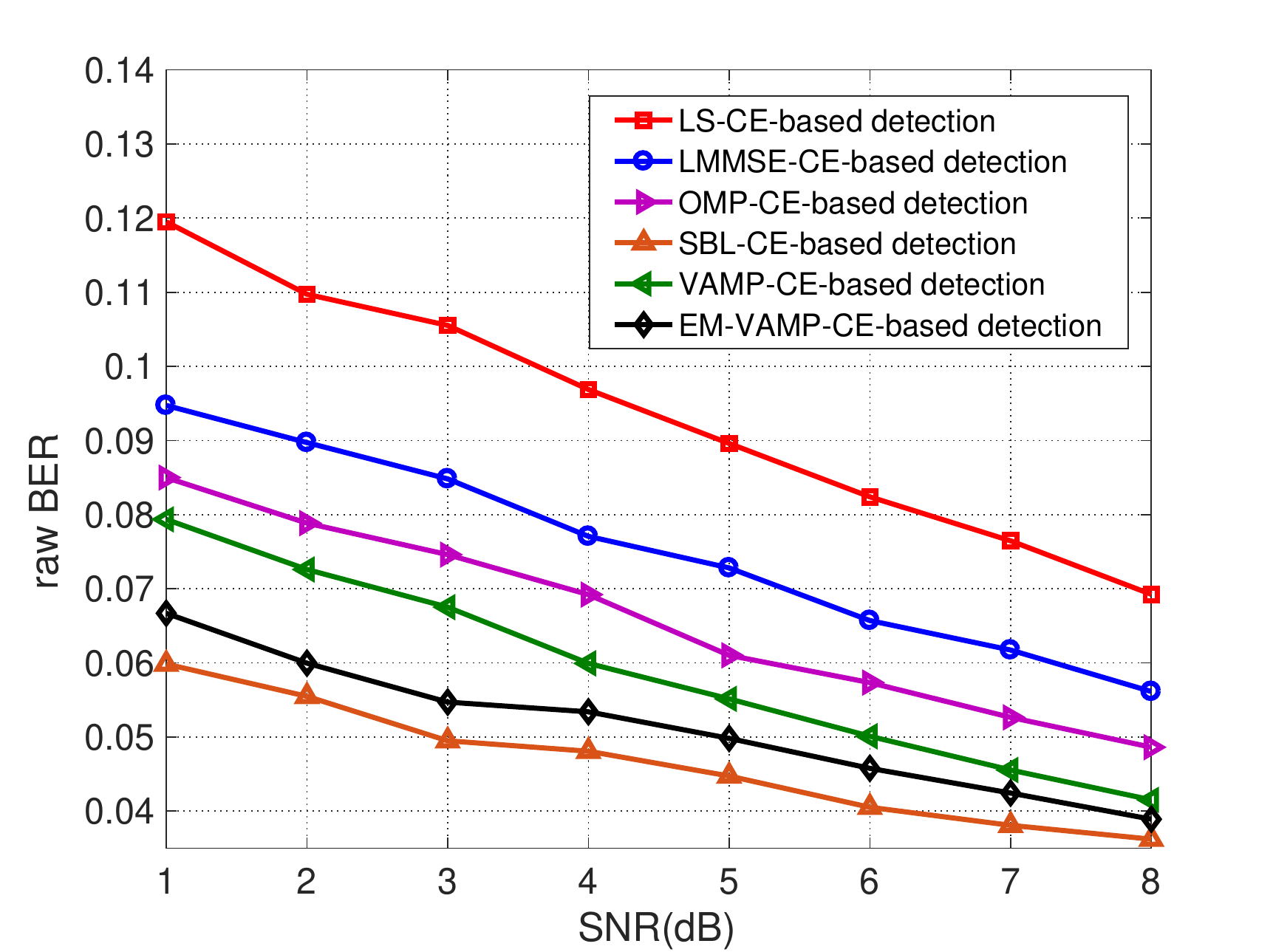}}
  \caption{The symbol MSE and raw BER comparisons at different SNRs (SPACE08 experiment).}
  \label{Sp08PerformanceComparison}
\end{figure}
The channel estimation results are next discussed. The comparison of symbol MSE and raw BER is shown in Fig.~\ref{Sp08PerformanceComparison}, where the number of used pilot symbols was 256. From the figure, the EM-VAMP-CE outperforms the LS, LMMSE, OMP, and VAMP CEs. Compared with the SBL-CE, it is inferior in performance but has a complexity that is three magnitude lower for this particular experiment.

\begin{figure}[!ht] %[!thbp]
   \hspace{-0.4cm}\centering
   \centering
   \includegraphics[width=13.5cm]
   {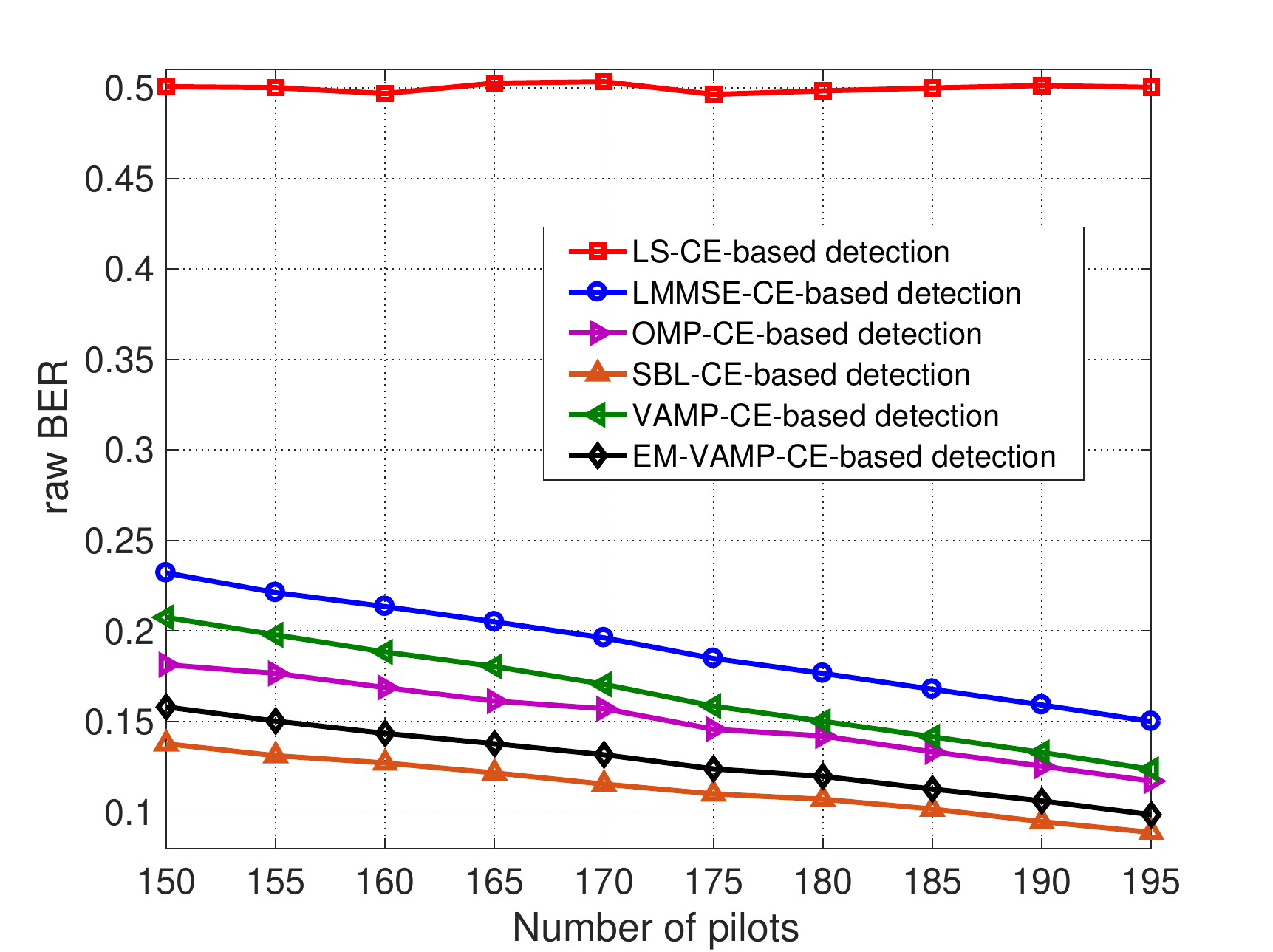}
   \caption{The raw BER comparison under different number of pilot symbols (SPACE08 experiment).}
   \label{Sp08PilotsNumber}
\end{figure}
For a MIMO transmission with $N$ transducers, the minimum number of pilots is $NL$ for the LS CE \cite{Tao18}.
%The performance as a function of pilot overhead is also investigated. We want to see how the reduction in the number of pilots affects each CE algorithm. According to \cite{Tao18}, the minimum number of pilots required by traditional channel estimation is $NL$. In this experiment, the minimum value is $K_p = 200$.
Therefore we focused on the case where the number of pilots is less than 200. The raw BER comparison among different CE methods at different number of pilots, is shown in Fig.~\ref{Sp08PilotsNumber}. Again, the EM-VAMP-CE achieves a performance only second to the SBL-CE.
%The number of pilots changed from 150 to 195. Raw BER is chosen as our indicator. The SNR is 5dB. Fig.~\ref{pilots number} shows that the EM-VAMP-CE consumes much less pilot resources to achieve the same performance.
%\begin{table}[!ht]
%%\small
%\renewcommand{\arraystretch}{0.5}
%\begin{center}
%\caption{\bf Performance comparison for different number of pilot symbols.}
%\label{hyperparameter estimation1}
%{\begin{tabular}{|c||p{1.5cm}<{\centering}|p{1.5cm}<{\centering}|p{1.5cm}<{\centering}|p{1.5cm}<{\centering}|p{1.5cm}<{\centering}|p{1.5cm}<{\centering}|p{1.5cm}<{\centering}|}\hline
%\diagbox{\bf Number of pilots}{${\bf Raw \enspace BER}$}{\bf CE Methods} & LS-CE & LMMSE-CE & OMP-CE & VAMP-CE & EM-VAMP-CE
%%\backslashbox
%\\\hline\hline $195$ & 0.0586 & 0.0295 & 0.0272& 0.0253 & 0.0245
%\\\hline $190$ & 0.0613 & 0.0319 & 0.0306 & 0.0258 & 0.0247
%\\\hline $185$ & 0.0630 & 0.0353 & 0.0337 & 0.0275 & 0.0254
%\\\hline $180$ & 0.0670 & 0.0382 & 0.0364 & 0.0297 & 0.0269\\\hline
%\end{tabular}}{}
%\end{center}
%\end{table}

\section{Conclusion}
\label{Conclusion}

In this paper, we proposed a new channel estimation scheme for MIMO OFDM underwater acoustic communications. It employed the VAMP algorithm to achieve approximate MMSE estimation of the channel. To determine the hyperparameters involved in the estimation, the EM algorithm was used. The proposed EM-VAMP channel estimation outperforms existing methods based on LS, LMMSE, OMP, as well as VAMP with predefined superparameters, according to simulation and experimental results. Even though it is slightly inferior to the SBL-CE in performance, it has a much lower complexity especially for a MIMO transmission employing non-orthogonal pilot sequences.%corroborated our claim. %based on data from the SPACE08 sea trial demonstrate the superiority of EM-VAMP estimator. To save computation complexity, the EM shares iterations with the VAMP.
%We established the mathematical model of MIMO channel estimation, and the derivation of EM-VAMP channel estimator is also completed. Then, we compare the performance of EM-VAMP-CE with LS, LMMSE, OMP and EM-GAMP estimator in terms of signal-to-noise ratio and occupying pilot resources.

\section*{Appendix A}
\label{AppendixA}

Plugging \eqref{VAMPestimator_1} into \eqref{VAMPestimator_4}, one gets
\begin{eqnarray}
\begin{aligned}
    \label{VAMPestimator_45}
&\hspace{-2cm}b_1(h_1(i)|r_{1,k}(i),\gamma_{1,k},{\bm \theta}_1) = \frac{\mathcal {CN}(h_1(i);r_{1,k}(i),{\gamma_{1,k}}^{-1})p(h_1(i);{\bm \theta}_1)}{\int{\mathcal {CN}( h_1(i);r_{1,k}(i),{\gamma_{1,k}}^{-1})} p(h_1(i);{\bm \theta}_1)dh_1(i)}\\
      &= \frac{\mathcal {CN}( h_1(i);r_{1,k}(i),{\gamma_{1,k}}^{-1})((1-\lambda)\delta(h_1(i))+\lambda{\mathcal {CN}(h_1(i);0,{\gamma_h}^{-1})})}{(1-\lambda)\mathcal {CN}(0;{r_{1,k}(i)},{\gamma_{1,k}}^{-1})+\lambda\mathcal {CN}(0;{r_{1,k}(i)},({\gamma_{1,k}}^{-1}+{\gamma_h}^{-1}))}\\
      &=\frac{(1-\lambda)\mathcal {CN}(0;r_{1,k}(i),{\gamma_{1,k}}^{-1})\delta(h_1(i))}{(1-\lambda)\mathcal {CN}(0;{r_{1,k}(i)},{\gamma_{1,k}}^{-1})+\lambda\mathcal {CN}(0;{r_{1,k}(i)},({\gamma_{1,k}}^{-1}+{\gamma_h}^{-1}))}+\\
      &\hspace{5mm}\frac{\lambda\mathcal {CN}(0;{r_{1,k}(i)},({\gamma_{1,k}}^{-1}+{\gamma_h}^{-1}))\mathcal{CN}({h_1(i);\mu_k(i),{\nu_k}})}{(1-\lambda)\mathcal {CN}(0;{r_{1,k}(i)},{\gamma_{1,k}}^{-1})+\lambda\mathcal {CN}(0;{r_{1,k}(i)},({\gamma_{1,k}}^{-1}+{\gamma_h}^{-1}))}\\
      &= (1-\pi_k(i)){\delta(h_1(i))}+{\pi_k(i)}{\mathcal {CN}({h_1(i);\mu_k(i),{\nu_k}})}
      %{\prod_{i=1}^{NL}}
\end{aligned}
\end{eqnarray}
where $\mu_k(i)$ and $\nu_k$ are defined in \eqref{VAMPestimator_7} and \eqref{VAMPestimator_8}.

\section*{Appendix B}
\label{AppendixB}
We take the first derivative of $Q_1({\bm \theta}_1,{{\bm \theta}_{1,k-1}})$ in \eqref{VAMPestimator_25} with respect to $\lambda$ and set it to zero, that is
%$\frac{\partial Q_1({\bm \theta}_1,{{\bm \theta}_{1,k-1}})}{\partial \lambda}$, set it to zero and get
\begin{align}
    \label{VAMPestimator_27}
    &\frac{\partial Q_1({\bm \theta}_1,{{\bm \theta}_{1,k-1}})}{\partial \lambda} = \frac{\partial}{\partial \lambda}\int{{\mathsf{ln}}(p({\uline{\bf h}}_m;{\bm \theta}_1)p({\bf y}_m|{\uline{\bf h}}_m;\gamma_w))b_1({\uline{\bf h}}_m|{\bf r}_{1,k},\gamma_{1,k},{\bm \theta}_{1,k-1})d{\uline{\bf h}}_m}\notag\\
    & = \frac{\partial}{\partial \lambda} \int({\mathsf{ln}} p({\uline{\bf h}}_m;{\bm \theta}_1)+{\mathsf{ln}}p({\bf y}_m|{\uline{\bf h}}_m;\gamma_w))b_1({\uline{\bf h}}_m|{\bf r}_{1,k},\gamma_{1,k},{\bm \theta}_{1,k-1})d{\uline{\bf h}}_m\notag\\
    & = \frac{\partial}{\partial \lambda}\int{{\mathsf{ln}} p({\uline{\bf h}}_m;{\bm \theta}_1)b_1({\uline{\bf h}}_m|{\bf r}_{1,k},\gamma_{1,k},{\bm \theta}_{1,k-1})d{\uline{\bf h}}_m}\notag\\
    & = \sum_{i=1}^{NL}\int b_1({\uline h}_m(i)|r_{1,k}(i),\gamma_{1,k},{\bm \theta}_{1,k-1})\frac{\partial}{\partial \lambda}{\mathsf{ln}}p({\uline h}_m(i);{\bm \theta}_1)d{\uline{h}}_m(i)=0.
\end{align}
For the BG prior probability in \eqref{VAMPestimator_1}, it is readily seen that
\begin{align}
    \frac{\partial}{\partial \lambda}{\mathsf{ln}}p({\uline h}_m(i);{\bm \theta}_1) &= \frac{{\mathcal {CN}}({\uline h}_m(i);0,{\gamma_h}^{-1})-\delta({\uline h}_m(i))}{p({\uline h}_m(i);{\bm \theta}_1)}\label{VAMPestimator_28}\\
    & = \left\{
                \begin{aligned}
                 &\frac{1}{\lambda}  &{\uline h}_m(i)\not=0 \label{VAMPestimator_29}\\
                 &\frac{-1}{1-\lambda}  &{\uline h}_m(i)=0
               \end{aligned}
               \right.
\end{align}
Substituting \eqref{VAMPestimator_29} into \eqref{VAMPestimator_27}, it becomes evident that the neighborhood around the point ${\uline h}_m(i)=0$ should be treated differently from the remainder of the complex space. Thus, we define the closed interval ${\mathcal B}_{\sigma}{\triangleq}[-\sigma,\sigma]$ and its complement ${\overline{\mathcal B}}_{\sigma}{\triangleq}{\mathcal{R}}\backslash{\mathcal B}_{\sigma}$, with $\mathcal{R}$ denoting the real space. In the limit $\sigma \rightarrow 0$, \eqref{VAMPestimator_27} can be transformed into
\begin{eqnarray}
\begin{aligned}
     \label{VAMPestimator_30}
     &\sum_{i=1}^{NL}\int_{\vert {\uline h}_m(i) \vert \in {\overline{\mathcal B}}_{\sigma}} b_1({\uline h}_m(i)|r_{1,k}(i),\gamma_{1,k},{\bm \theta}_{1,k-1})d{\uline h}_m(i)\\
     &= \frac{\lambda}{1-\lambda}\sum_{i=1}^{NL}\int_{\vert {\uline h}_m(i) \vert \in {\mathcal B}_{\sigma}} b_1({\uline h}_m(i)|r_{1,k}(i),\gamma_{1,k},{\bm \theta}_{1,k-1})d{\uline h}_m(i).
\end{aligned}
\end{eqnarray}
According to \eqref{VAMPestimator_5}, \eqref{VAMPestimator_30} is equivalent to
\begin{eqnarray}
     \label{VAMPestimator_31}
     \sum_{i=1}^{NL}{\pi_k(i)}= \frac{\lambda}{1-\lambda}\sum_{i=1}^{NL}(1-\pi_k(i)),
\end{eqnarray}
based on which the optimal estimation of $\lambda$ at the $k$-th iteration is solved as
\begin{eqnarray}
     \label{VAMPestimator_32}
      \lambda_{k} = {\frac{1}{NL}}\sum_{i=1}^{NL}\pi_k(i).
\end{eqnarray}

Similarly, we take the first derivative of $ \frac{\partial Q_1({\bm \theta}_1,{{\bm \theta}_{1,k-1}})}{\partial {\gamma_h}}$ with respect to ${\gamma_h}$ and set it to zero, leading to
%perform operations similar to \eqref{VAMPestimator_27} and finally get
\begin{align}
\label{VAMPestimator_33}
\frac{\partial Q_1({\bm \theta}_1,{{\bm \theta}_{1,k-1}})}{\partial {\gamma_h}} = \sum_{i=1}^{NL}\int b_1({\uline h}_m(i)|r_{1,k}(i),\gamma_{1,k},{\bm \theta}_{1,k-1})\frac{\partial}{\partial {\gamma_h}}{\mathsf{ln}}p({\uline h}_m(i);{\bm \theta}_1)d{\uline h}_m(i)=0.
\end{align}
For the BG prior probability in \eqref{VAMPestimator_1}, it is readily seen that
\begin{align}
\frac{\partial}{\partial {\gamma_h}}{\mathsf{ln}}p({\uline h}_m(i);{\bm \theta}_1)d{\uline h}_m(i) &= (\frac{1}{\gamma_h}-{\vert {\uline h}_m(i) \vert}^2)\frac{\lambda{\mathcal {CN}}({\uline h}_m(i);0,{\gamma_h}^{-1})}{p({\uline h}_m(i);{\bm \theta}_1)}\label{VAMPestimator_34}\\
& = \left\{
                \begin{aligned}
                 &\frac{1}{\gamma_h}-{\vert {\uline h}_m(i) \vert}^2  &{\uline h}_m(i)\not=0 \label{VAMPestimator_35}\\
                 &0  &{\uline h}_m(i)=0
               \end{aligned}
               \right.
\end{align}
Dividing the integration domain in \eqref{VAMPestimator_33} into ${\mathcal B}_{\sigma}$ and ${\overline{\mathcal B}}_{\sigma}$ as before
then plugging \eqref{VAMPestimator_35}, \eqref{VAMPestimator_33} is equivalent to the following in the limit $\sigma \rightarrow 0$
\begin{eqnarray}
\label{VAMPestimator_36}
\sum_{i=1}^{NL}\int_{\vert {\uline h}_m(i)\vert \in {\overline{\mathcal B}}_{\sigma}}(1-{\gamma_h}{\vert {\uline h}_m(i) \vert}^2)b_1({\uline h}_m(i)|r_{1,k}(i),\gamma_{1,k},{\bm \theta}_{1,k-1})d{\uline h}_m(i)=0,
\end{eqnarray}
based on which the optimal $\gamma_h$ is solved as
\begin{eqnarray}
\begin{aligned}
     \label{VAMPestimator_37}
      \gamma_{h,k} &= \frac{\sum_{i=1}^{NL}{\mathsf{lim}}_{\sigma \rightarrow 0}\int_{\vert {\uline h}_m(i)\vert \in {\overline{\mathcal B}}_{\sigma}}b_1({\uline h}_m(i)|r_{1,k}(i),\gamma_{1,k},{\bm \theta}_{1,k-1})d{\uline h}_m(i)}{\sum_{i=1}^{NL}{\mathsf{lim}}_{\sigma \rightarrow 0}\int_{\vert {\uline h}_m(i)\vert \in {\overline{\mathcal B}}_{\sigma}}{\vert {\uline h}_m(i)\vert}^2b_1({\uline h}_m(i)|r_{1,k}(i),\gamma_{1,k},{\bm \theta}_{1,k-1})d{\uline h}_m(i)}\\
      &= \big({\frac{1}{\lambda_{k} NL}}\sum_{i=1}^{NL}\pi_k(i)({\vert\mu_k(i)\vert^2}+\nu_k)\big)^{-1}
\end{aligned}
\end{eqnarray}
where the equality of \eqref{VAMPestimator_32} has been used.

\section*{Appendix C}
\label{AppendixC}
We take the first derivative of $\frac{\partial Q_2(\gamma_w,\gamma_{w,k-1})}{\partial {\gamma_w}}$ in \eqref{VAMPestimator_38} with respect to ${\gamma_w}$ and set it to zero, obtaining
\begin{eqnarray}
\begin{aligned}
    \label{VAMPestimator_39}
    &\frac{\partial Q_2(\gamma_w,\gamma_{w,k-1} )}{\partial \gamma_w} = \frac{\partial}{\partial \gamma_w}\int{\mathsf{ln}}(p({\uline{\bf h}}_m;{\bm \theta}_1)p({\bf y}_m|{\uline{\bf h}}_m;\gamma_w))b_2({\bf h}_m|{\bf r}_{2,k},\gamma_{2,k},\gamma_{w,k-1})d{\uline{\bf h}}_m\\
    & = \frac{\partial}{\partial \gamma_w} \int({\mathsf{ln}}p({\uline{\bf h}}_m;{\bm \theta}_1)+{\mathsf{ln}}p({\bf y}_m|{\uline{\bf h}}_m;\gamma_w)))b_2({\uline{\bf h}}_m|{\bf r}_{2,k},\gamma_{2,k},\gamma_{w,k-1})d{\uline{\bf h}}_m\\
    & = \frac{\partial}{\partial \gamma_w}\int{{\mathsf{ln}}(p({\bf y}_m|{\uline{\bf h}}_m;\gamma_w)))b_2({\uline{\bf h}}_m|{\bf r}_{2,k},\gamma_{2,k},\gamma_{w,k-1})d{\uline{\bf h}}_m}\\
    & = \int b_2({\uline{\bf h}}_m|{\bf r}_{2,k},\gamma_{2,k},\gamma_{w,k-1})\frac{\partial}{\partial \gamma_w}{\mathsf{ln}}p({\bf y}_m|{\uline{\bf h}}_m;\gamma_w)d{\uline{\bf h}}_m\\
    & = 0.
\end{aligned}
\end{eqnarray}
It is easy to have
\begin{align}
\frac{\partial}{\partial \gamma_w}{\mathsf{ln}}p({\bf y}_m|{\uline{\bf h}}_m;\gamma_w) &= \frac{\partial}{\partial \gamma_w}{{\mathsf{ln}}({(\frac{\gamma_w}{\pi})}^{K_p}{\mathsf{exp}}(-\gamma_w{\Vert{\bf y}_m-{\bf W}{\uline{\bf h}}_m \Vert}_2^2))}\notag\\
& = \frac{K_p}{\gamma_w}-{\Vert  {\bf y}_m-{\bf W}{\uline{\bf h}}_m \Vert}_2^2.\label{VAMPestimator_40}
\end{align}
Plugging \eqref{VAMPestimator_40} into \eqref{VAMPestimator_39}, we solve $\gamma_w$ as
\begin{align}
     &\gamma_{w,k} = \bigg(\frac{1}{K_p}\int{\Vert {\bf y}_m-{\bf W}{\uline{\bf h}}_m \Vert}_2^2)b_2({\uline{\bf h}}_m|{\bf r}_{2,k},\gamma_{2,k},\gamma_{w,k-1})d{\uline{\bf h}}_m\bigg)^{-1}\\
     &=\bigg(\frac{1}{K_p}[{\Vert {\bf y}_m - {\bf W}{\widehat{\bf h}}_{2,k} \Vert}^{2}+\text{Tr}({\bf W}(\gamma_{w,k-1}{\bf W}^{H}{\bf W}+\gamma_{2,k}{\bf I}){\bf W}^{H})]\bigg)^{-1}\label{VAMPestimator_41}\\
     & = \bigg(\frac{1}{K_p}{\Vert {\bf y}_m - {\bf W}{\widehat{\bf h}}_{2,k} \Vert}^{2}+\frac{1}{K_p}{\gamma^{-1}_{w,k-1}}\sum_{n=1}^{R}(\frac{{\vert s_n \vert}^{2}}{{\vert s_n \vert}^{2}+{\gamma^{-1}_{w,k-1}}{\vert s_n \vert}^{2}})\bigg)^{-1}\label{VAMPestimator_42}
\end{align}
where \eqref{VAMPestimator_41} has used \eqref{VAMPestimator_16}. The \eqref{VAMPestimator_42} is obtained by replacing ${\bf W}$ in \eqref{VAMPestimator_41} by its SVD ${\bf W} = {\bf U}{\bf S}{\bf V}^{H}$ for reduced complexity. %Its purpose is to simplify the form of expression.

\bibliographystyle{IEEEtran}
\normalem
\bibliography{myreferences}
\end{document}